\documentstyle[mnras_cite,psfig]{mn}

\def\gsim{~\rlap{$>$}{\lower 1.0ex\hbox{$\sim$}}}
\def\lsim{~\rlap{$<$}{\lower 1.0ex\hbox{$\sim$}}}
\def\d{{\rm d}}
\def\c{{\rm c}}

\def\G{{\rm G}}
\def\kb{{\rm k_B}}
\def\mh{{\rm m_H}}

\def\e{e}
\def\HI{H{\sc i}}
\def\HII{H{\sc ii}}
\def\HeI{He{\sc i}}
\def\HeII{He{\sc ii}}
\def\HeIII{He{\sc iii}}
\def\TIGM{T_{\rm IGM}}

\begin{document}

\title{The Effects of Photoionization on Galaxy Formation --- I: Model and Results at z=0} 
\author[A.~J.~Benson, C.~G.~Lacey, C.~M.~Baugh, S.~Cole,\&
C.~S.~Frenk]{A.~J.~Benson$^1$,  C.~G.~Lacey$^2$ C.~M.~Baugh$^3$,
S.~Cole$^3$, \& C.~S.~Frenk$^3$ \\
1. California Institute of Technology, MC 105-24, Pasadena, CA 91125, U.S.A. (e-mail: abenson@astro.caltech.edu) \\
2. SISSA, Astrophysics Sector, via Beirut 2-4, 34014 Trieste, Italy\\
3. Physics Department, University of Durham, Durham, DH1 3LE, England} 

\maketitle

\begin{abstract}
We develop a coupled model for the evolution of the global properties
of the intergalactic medium (IGM) and the formation of galaxies, in
the presence of a photoionizing background due to stars and quasars.
We use this model to predict the thermodynamic history of the IGM when
photoionized by galaxies forming in a cold dark matter (CDM) universe.
The evolution of the galaxies is calculated using a semi-analytical
model, including a detailed treatment of the effects of tidal
stripping and dynamical friction on satellite galaxies orbiting inside
larger dark matter halos. We include in the model the negative
feedback on galaxy formation from the photoionizing
background. Photoionization inhibits galaxy formation in low-mass dark
matter halos in two ways: (i) heating of the IGM and inhibition of the
collapse of gas into dark halos by the IGM pressure, and (ii)
reduction in the radiative cooling of gas within halos. The result of
our method is a self-consistent model of galaxy formation and the
IGM. The IGM is reheated twice (during reionization of \HI\ and
\HeII), and we find that the star formation rate per unit volume is
slightly suppressed after each episode of reheating. We find that
galaxies brighter than $L_{\star}$ are mostly unaffected by
reionization, while the abundance of faint galaxies is significantly
reduced, leading to present-day galaxy luminosity functions with
shallow faint end slopes, in good agreement with recent observational
data. Reionization also affects other properties of these faint
galaxies, in a readily understandable way.\\

\end{abstract}

\section{Introduction}

It is now known that the hydrogen in the intergalactic medium (IGM),
which became neutral at $z\sim 1000$ \cite{peebles68,zeldovich68},
must have been reionized somewhere between redshifts 6 and 30, the
lower limit coming from the lack of a Gunn-Peterson trough in quasar
spectra at that redshift (e.g. \pcite{fan00}), and the upper limit
from the bound on the optical depth to the last scattering surface
measured from the cosmic microwave background (CMB;
\pcite{nett01}). In fact, very recent results
\cite{djorgovski01,becker01} suggest reionization very close to the
lower limit of this range. If there are large populations of galaxies
or quasars at high redshifts, as is predicted by current structure
formation models (e.g. \pcite{benson00c}) and as confirmed up to
redshifts $\approx 6$ observationally \cite{fan00,stern00}, then
reionization is most likely to have occurred through photoionization,
as both galaxies and quasars emit copious quantities of ionizing
photons (e.g. \pcite{cr86}). Several models of reionization have been
developed in recent years
\cite{haimanloeb96,gnedinostriker97,chiu99,valageas99,ciardi00,gnedin2000a,miralda00,benson00c},
many reaching the conclusion that reionization occurred between
$z\approx$7--12, although large systematic uncertainties remain due to
uncertainties in the efficiency of galaxy formation, in the fraction
of ionizing photons that escape a galaxy and in the density
distribution of ionized gas in the IGM (see, for example,
\pcite{benson00c}). If this picture of reionization is correct then it
is clear that the thermodynamic history of the IGM is determined by
the formation and evolution of galaxies and quasars.

The photoionizing background responsible for reionizing the IGM may
also act, directly and indirectly, to inhibit galaxy formation, as was
first pointed out by \scite{dor67}, and first investigated in the
context of CDM models by \scite{cr86}.  Galaxies are thought to form
by a two-stage collapse process, in which gas first collapses into
dark matter halos along with the dark matter itself, and then
collapses relative to the dark matter within halos if it is able to
cool radiatively to below the halo virial temperature, thus losing its
pressure support. The second stage of the collapse is necessary in
order to increase the gas density to the point where it becomes
self-gravitating relative to the dark matter, which is believed to be
a necessary condition for the gas to be able to fragment to form
stars. In the presence of an ionizing background, both stages of this
collapse process are inhibited, particularly for low mass
halos. Firstly, the ionizing background heats the IGM to temperatures
of around $10^4$K, and the resulting thermal pressure of the gas then
prevents it from collapsing into low mass halos along with the dark
matter. Secondly, the ionizing background reduces the rate of
radiative cooling of gas inside halos, mainly by reducing the
abundance of neutral atoms which can be collisionally excited. Both of
these mechanisms will strongly inhibit galaxy formation in halos with
virial temperatures less than $\sim 10^4$K, and so may have important
effects on the faint end of the galaxy luminosity function and also on
the properties of the dwarf satellite galaxies of the Milky Way and
other galaxies.

There have been many studies of the effects of an ionizing background
on galaxy formation, both analytical
(e.g. \pcite{efstath92,babul92,chiba94,thoul96,kepner97,nagashima99})
and using numerical simulations
(e.g. \pcite{vedel94,quinn96,weinberg97,navarro97}), but in most of
these the ionizing background was simply taken as an external input. A
few studies have investigated the more difficult self-consistent
problem, relating the ionizing background at any redshift to the
fraction of baryons which had previously collapsed to form galaxies,
and at the same time including the effect of the ionizing background
in inhibiting further galaxy formation
(e.g. \pcite{shapiro94,gnedinostriker97,valageas99}). The analytical
studies have used a wide variety of approaches and approximations, but
have generally modelled galaxy formation and the effects of
photoionization only in a very simplified or partial way (e.g. for
photoionization, either considering only the supression of collapse
into dark halos, or the suppression of cooling within dark halos). On
the other hand, the numerical studies were limited in the predictions
they could make about properties of the present-day galaxy population
by the range of physics included and by their limited dynamical
range. In the present paper, we present a new model for the coupled
evolution of the IGM, the ionizing background and galaxies, based on a
semi-analytical model of galaxy formation, enabling us to determine in
much more detail than in previous studies the effects of
photoionization on observable galaxy properties. Compared to previous
analytical studies (in particular \pcite{valageas99}), the main
improvements are that we have a much more detailed model for galaxy
formation through hierarchical clustering, including many different
processes, and a more accurate model for how photoionization
suppresses galaxy formation through the two mechanisms described
above. In particular, the suppression of gas collapse into dark matter
halos due to the IGM pressure is modelled based on the latest results
from gas-dynamical simulations.

Our starting point is the semi-analytic model of galaxy formation
described by \scite{cole2000}, which attempts to model the galaxy
formation process {\it ab initio}, in the framework of structure
formation through hierarchical clustering. We then modify this to
include the new physics we are interested in here. We develop a model
for the evolution of the thermodynamic properties of the IGM in the
presence of the ionizing radiation background produced by galaxies and
quasars, the former predicted by the semi-analytic model, and the
latter based on observational data.  We are then able to predict the
mean temperature of the IGM and the spectrum of the ionizing
background as functions of cosmic time. We adapt the Cole et al. model
to determine the mass of gas able to accrete onto each dark matter
halo from the heated IGM, and to include the effects of heating by the
ionizing background. Finally, we include a more detailed treatment of
the dynamical evolution of satellites orbiting within larger dark
matter halos, including the effects of tidal stripping. The approach
of investigating the effects of photoionization on galaxy formation by
using a semi-analytic model was previously used by
\scite{nagashima99}, but they considered only the heating of gas in
halos by the UV background, and so our current work represents a more
thorough treatment of the problem, as well as being based on a
much-improved galaxy formation model.

There are two parts to this paper. Firstly, we describe how the
physics of the IGM/galaxy interaction may be modelled in a simple
way. Secondly, we present results from our model, focussing on the
evolution of the IGM and ionizing background and the properties of
local galaxy population. We briefly comment on how high redshift
galaxies are affected. In a companion paper \cite{benson01b}, we will
explore in detail the consequences of our model for the population of
satellite galaxies seen in the Local Group.

The remainder of this paper is organized as follows. In
\S\ref{sec:model} and \S\ref{sec:tiddis} we describe in detail our
model for the evolution of the IGM and galaxy formation. In
\S\ref{sec:results} we present results from this model for the
evolution of the IGM and the population of galaxies at the present day
in the currently favoured $\Lambda$CDM model of structure
formation. Finally, in \S\ref{sec:discuss} we present our conclusions.

\section{Model of phoionization and IGM evolution}
\label{sec:model}

We use the semi-analytic model of galaxy formation developed by
\scite{cole2000} to determine the properties of galaxies in the
Universe. The model includes formation and merging of dark matter
halos, shock-heating and radiative cooling of gas within halos,
collapse of cold gas to form galaxy disks, star formation from the
cold gas, galaxy mergers within common dark matter halos leading to
formation of galaxy spheroids, chemical enrichment, and the luminosity
evolution of stellar populations.  The fiducial model of
\scite{cole2000} (for which $\Omega_0 = 0.3$, $\Lambda_0=0.7$,
$\Omega_{\rm b}=0.02$ and $h=0.7$
\footnote{We define Hubble's constant to be $H_0=100 h\ {\rm km s^{-1}
Mpc^{-1}}$.}) has been shown to reproduce many of the properties of
galaxies in the local Universe, such as their luminosity functions,
the slope and scatter of the Tully-Fisher relation, colours, sizes and
metallicities \cite{cole2000} and also the clustering of galaxies in
real and redshift space \cite{benson00dummy}.

The model of \scite{cole2000}, like most other semi-analytic models of
galaxy formation (e.g. \pcite{kauffmann93,somerville99}), includes a
prescription for feedback due to energy input from supernovae and
stellar winds. This is assumed to reheat cold gas and eject it from
galaxies, thus inhibiting galaxy formation in low mass dark matter
halos. Several studies of how this feedback may physically operate can
be found in the literature \cite{dekel87,maclow99,goodwin2000}. This
feedback is required in CDM models in order to produce a faint-end
slope of the local galaxy luminosity function which is as shallow as
that observed \cite{wr78,cole91,wf91}, and also to produce galactic
disks of sizes comparable to those observed \cite{cole2000}.

While ejection of gas by supernovae driven outflows is undoubtedly an
important process (e.g. \pcite{martin99}), other processes may also
inhibit galaxy formation, for example, preheating of the IGM
\cite{blanchard92,evrard91,kaiser91,valageas99}, heating of the gas
inside galaxy and cluster halos \cite{wu00,bower2000}, and the effects
of a photoionizing background.  The last of these is perhaps the best
studied (see, for example,
\pcite{efstath92,thoul96,katz96,bullock00}). A photoionizing
background both supplies heat to the gas through ionization, and
reduces the rate at which the gas can cool by reducing the abundance
of neutral atomic species which can be collisionally excited. It thus
raises the IGM temperature and so prevents it from collapsing into
small halos, and also reduces the cooling rate of gas within halos and
so reduces the fraction of baryons which can collapse to form a
galaxy. For the formation of the very first objects at high redshift,
cooling of the gas by molecular hydrogen is probably important, and
one needs to consider the dissociation of these molecules by
non-ionizing UV radiation (e.g. \pcite{ciardi00}), but these processes
are only important well before the epoch of reionization in our model,
since conversion of only a very tiny fraction of the baryons into stars
is sufficient to produce enough UV radiation to dissociate all of the
$H_2$ molecules.

In this section, we describe how we modify the model of
\scite{cole2000} to calculate the evolution of the IGM temperature and
ionizing background, the suppression of gas collapse into halos by the
IGM pressure, and the suppression of cooling within halos by the
ionizing background. Our modelling of the dynamical evolution of
satellite galaxies within larger halos is described in \S\ref{sec:tiddis}.

\subsection{Evolution of the Ionizing Background and the IGM Temperature}

We will treat the IGM as a mixture of six species (\e, \HI, \HII,
\HeI, \HeII\ and \HeIII) which interact with each other and with a
uniform background of radiation emitted by stars and quasars. Since we
are here primarily interested in the properties of low-redshift galaxies,
we will not include H$_2$ in our calculations, since it will be
dissociated at high redshifts (e.g. \pcite{ciardi00}). We follow the
evolution of the abundances of these species and the gas temperature
for parcels of gas spanning a wide range in density contrast. The
density contrast of each parcel is allowed to change with time as
described in \S\ref{sec:pdf}. Here, we treat all gas in the Universe
as being part of the IGM. Since the fraction of the total gas content
of the Universe which becomes part of a galaxy in our model is always
small this is a reasonable approximation. Some gas should of course
fall into the potential wells of dark matter halos (see
\S\ref{sec:mfilter}). Since this gas typically occupies a small
fraction of the volume of the Universe we ignore it for calculating
the properties of the IGM.

In the remainder of this section we describe in detail how we model
the evolution of the ionizing background and IGM temperature.

\subsubsection{Evolution of Gas Density}
\label{sec:pdf}

We wish to calculate the thermodynamic behaviour of gas in the IGM up
until the point at which it falls into a virialised dark matter
halo. The gas in the IGM will have a range of overdensities resulting
from the growth of density fluctuations due to gravitational
instability (we do not consider here the possibility of a multiphase
medium which may also produce variations in gas density). Since
recombination rates, and consequently heating and cooling rates,
depend on the gas density, it is necessary to take this evolving
distribution of densities into account in our model.

We characterise the evolving distribution of gas densities via the
probability distribution function (PDF), $P_{\rm V}(\Delta,t)$,
defined such that $P_{\rm V}(\Delta,t)\d \Delta$ is the fraction of
volume in the universe occupied by gas with a density contrast
$\Delta=\rho/\overline{\rho}$ at time $t$, where $\rho$ is the gas
density at a point and $\overline{\rho}$ is the mean gas density in
the Universe.  Normalisation of this function to give the correct mean
density and total mass requires that
\begin{equation}
\int_0^\infty P_{\rm V}(\Delta,t)\, \d \Delta = 1,
\label{eq:norm1}
\end{equation}
and
\begin{equation}
\int_0^\infty \Delta\, P_{\rm V}(\Delta,t)\, \d \Delta = 1.
\label{eq:norm2}
\end{equation}
The fraction of mass with density contrast $\leq \Delta$ is given by
\begin{equation}
F(\Delta,t) = \int_0^{\Delta} \Delta^\prime P_{\rm V}(\Delta^\prime,t)\, \d \Delta^\prime .
\end{equation}
We assume that as the gas density field evolves, the ranking of gas
elements by density remains the same. The density contrast at time $t$ of a
gas element which has density contrast $\Delta_0$ at time $t_0$ is
therefore given by the solution of
\begin{equation}
F(\Delta[t],t)=F(\Delta_0,t_0),
\label{eq:deltaevol}
\end{equation}

We can use eqn.~(\ref{eq:deltaevol}) to calculate the evolution in
overdensity $\Delta[t]$ of individual parcels of IGM gas having
different values of $\Delta_0$, once the functional form and evolution
of $P_{\rm V}(\Delta,t)$ have been specified.  In our standard model,
we assume that the PDF has a log-normal form, which has been found to
provide a reasonable description of the density distribution produced
by gravitational instability in the mildly non-linear regime
(e.g. \pcite{coles91}),
\begin{equation}
P_{\rm V}(\Delta) = \left({A\over \Delta}\right) \exp\left[ {(\ln \Delta - \overline{\ln \Delta} )^2
\over 2 \sigma_\Delta^2 } \right],
\end{equation}
Here, $\sigma_\Delta$ determines
the width of the distribution and the constants $A$ and $\overline{\ln
\Delta}$ are fixed from the normalisation conditions
(eqns.~\ref{eq:norm1} and \ref{eq:norm2}). The value of
$\sigma_\Delta$ as a function of time can be chosen to reproduce a
desired baryonic clumping factor
\begin{equation}
f_{\rm clump} \equiv \frac{\overline{\rho^2}}{{\overline\rho}^2}
= \int_0^\infty \Delta^2\, P_{\rm V}(\Delta,t)\, \d \Delta,
\end{equation}
where the overbar denotes a volume average.  In particular, we will
choose $\sigma_\Delta$ to reproduce the baryonic clumping factor,
$f_{\rm clump}^{\rm (variance)}$, derived by \scite{benson00c}. In
their calculation, \scite{benson00c} assumed that gas in the IGM
essentially traces the dark matter except that pressure prevents the
gas from falling into dark matter halos with virial temperatures less
than $10^4$K. They then calculated $f_{\rm clump}^{\rm
(variance)}=1+\sigma^2$, where $\sigma^2$ is the variance of the dark
matter density field in spheres of radius equal to the radius of a
$10^4$K halo ($\sigma^2$ was calculated from the smoothed non-linear
dark matter power spectrum obtained using the techniques of
\pcite{pd96}). In the present work, the halo mass below which gas
accretion is negligible varies as a function of time. Nevertheless,
our estimate of the clumping factor should still provide a reasonable
approximation.  We note that at the redshift appropriate for \HI\
reionization in this work (see \S\ref{sec:IGMprops}) the two different
clumping factors considered by \scite{benson00c} are in fact very
similar (see their Fig.~9).

Once the evolution of the clumping factor has been chosen, our model
results are insensitive to the particular functional form chosen for
the PDF. For example, if instead of the lognormal distribution we use
the form 
\begin{equation}
P_{\rm V}(\Delta) = \left({A\over \Delta}\right) \exp\left[ {(|\ln \Delta - \overline{\ln \Delta}| )^3
\over 2 \sigma_\Delta^3 } \right],
\end{equation}
which falls off much more rapidly away from $\ln \Delta =
\overline{\ln \Delta}$, this makes negligible difference to the
evolution of the mean IGM temperature, ionization state and the
spectrum of the ionizing background. We truncate the distribution of
gas densities above $\Delta = 300$, which is roughly the mean density
contrast of halos at $z=0$ in our adopted cosmology, because reaction
rates become extremely rapid for higher densities, making solution of
the rate equations numerically difficult. Gas at higher overdensities
accounts for only a small fraction of the total volume, and we have
checked that moving the truncation point to larger $\Delta$ makes
little difference to our results.

\subsubsection{Background Radiation}

We follow the proper number density of photons per unit frequency,
$n_\nu$, which evolves with time as
\begin{eqnarray}
{\partial n_\nu \over \partial t} & = & {\dot{a}\over a} \left[ -3 n_\nu +  
\frac{\partial}{\partial\nu} \left( \nu n_\nu \right)  \right] 
+ S_\nu \nonumber \\
 & &  - \sum_i \sum_j \c \sigma_{\nu,i} f_{{\rm v},j} n_{i,j} n_\nu ,
\label{eq:nnu}
\end{eqnarray}
where $\c$ is the speed of light, the term $-3 (\dot{a}/a) n_\nu$ on
the right-hand side represents the dilution of the number density by
the Hubble expansion, and the term $(\dot{a}/a) \partial \left( \nu
n_\nu \right)/\partial\nu$ describes the effect of the redshifting of
the photon frequencies.  Here $S_\nu$ is the emissivity (i.e. number
of photons emitted per unit volume, per unit time, per unit
frequency), $\sigma_{\nu,i}$ is the photoionization cross section for
species $i$ (\HI, \HeI, \HeII), $f_{{\rm v},j}$ is the fraction of the
volume of the universe occupied by gas in density bin $j$, and
$n_{i,j}$ is the abundance of species $i$ in density bin $j$.

The photon number density is related to the background intensity by
\begin{equation}
J_\nu = {\c {\rm h_P} \nu \over 4 \pi} n_\nu,
\end{equation}
where $J_\nu$ is the intensity per unit solid angle per unit frequency
and ${\rm h_P}$ is Planck's constant.

\subsubsection{Rate Equations}

The evolution of the abundances of the different ionization states of
H and He is described by equations of the form
\begin{eqnarray}
{\d n_i \over \d t} & = & \left[ \alpha_i(\TIGM)n_{i+1} n_{\rm e} - \alpha_{i-1}(\TIGM)n_i n_{\rm e} \right. \nonumber \\
 & & - \Gamma_{{\rm e},i}(\TIGM) n_i n_{\rm e} + \Gamma_{{\rm e},i-1}(\TIGM) n_{i-1} n_{\rm e} \nonumber \\ 
 & & \left. - \Gamma_{\gamma,i} n_i + \Gamma_{\gamma,i-1} n_{i-1} \right] \nonumber \\
 & &  + \left( {1 \over \Delta(a)} {\d \Delta(a)\over \d t} - 3 {\dot{a}\over a}\right) n_i,
\end{eqnarray}
where for each atomic species H or He, $i$ refers to the ionization
state (i.e. $i=1,2$ for \HI, \HII\ and $i=3,4,5$ for \HeI, \HeII,
\HeIII), $n_i$ is the proper number density, $\TIGM$ is the
temperature, $\alpha_i$ is the recombination rate coefficient to $i$,
$\Gamma_{{\rm e},i}$ is the collisional ionization rate coefficient
from $i$ and $\Gamma_{\gamma,i}$ is the photoionization rate
coefficient from $i$.  The evolution of the electron density then
follows from the conservation of the total number of electrons.

We consider the evolution of a parcel of gas of density contrast
$\Delta(t)$, which has a thermal energy per unit volume given by
$E=\frac{3}{2}{\rm k_B}\TIGM n_{\rm tot}$, where $n_{\rm tot}$ is the
total number of particles per unit volume. The energy changes due to
adiabatic expansion/compression and atomic heating/cooling
processes. Thus the evolution of $E$ may be written as
\begin{equation}
{\d E \over \d t} = {5 \over 3} \left(  {1 \over \Delta(a)} {\d
\Delta(a)\over \d t} - 3 {\dot{a}\over a} \right) E +
\left(\Sigma^{\rm T} - \Lambda^{\rm T}\right),
\label{eq:Eevol}
\end{equation}
where the first term represents adiabatic expansion or compression and
the second represents atomic heating and cooling
processes. $\Sigma^{\rm T}$ is the rate of heating per unit volume due
to all heat sources (i.e. photoionization and Compton heating) and
$\Lambda^{\rm T}$ is the rate of cooling per unit volume due to all
heat sinks (i.e. Compton cooling and various atomic processes). We use
the notation $\Sigma^{\rm T}$ and $\Lambda^{\rm T}$ to indicate rates
of thermal energy gain/loss, as distinct from the usual radiative
cooling function $\Lambda$ which includes the entire energy of the
photons emitted by recombinations. Note that the evolution of the gas
density is entirely determined by the functional form of the PDF
$P_{\rm V}(\Delta)$ and the redshift evolution of $f_{\rm clump}$ and
is unaffected by any heating/cooling of the gas. In reality, the gas
density distribution should respond to differences in gas
pressure. However, the effects of pressure forces should be important
only on scales smaller than the Jeans length (or, more precisely, the
filtering length to be introduced in the next subsection), which
always remains small ($\lsim 1h^{-1}$Mpc) relative to the much larger
scales over which we calculate volume averages.

From eqn.~(\ref{eq:Eevol}) and the definition of $E$, we obtain the
following equation for the evolution of the IGM temperature:
\begin{eqnarray}
{1 \over \TIGM} {\d \TIGM \over \d t} & = & -2 {\dot{a} \over a} + {2 \over 3 \Delta} {\d \Delta \over \d t} + {\Sigma^{\rm T} - \Lambda^{\rm T} \over \frac{3}{2}{\rm k_B}\TIGM n_{\rm tot}} \nonumber \\
 &  & - {\Delta \over a^3 n_{\rm tot}} {\d  \over \d t}
\left({a^3 n_{\rm tot} \over \Delta}\right).
\label{eq:TIGMevol}
\end{eqnarray}
The final term accounts for the effects of changes in the total
particle number density due to ionization/recombination.  For a
homogeneous IGM ($\Delta=1$) with no heating or cooling and no
ionization or recombination, we have simply $n_{\rm tot}\propto
a^{-3}$ and $T_{\rm IGM}\propto a^{-2}$.

Equations~\ref{eq:nnu}--\ref{eq:TIGMevol} describe the evolution of a
parcel of gas of specified final density contrast $\Delta_0$. These
equations, along with those describing the evolution of the background
radiation spectrum, are solved for a range of $\Delta_0$ using a
modified Bulirsch-Stoer method which is applicable to this stiff set
of equations \cite{bader83}. The matrix decomposition that must be
carried out as part of this method is efficiently achieved using a
suitable sparse matrix package.

The initial conditions for the abundance of each species and for the
temperature are taken from the {\sc recfast} code \cite{seager2000}
which accurately evolves the IGM through the recombination epoch (we
typically begin our own calculations at $z=200$, at which point
recombination is essentially complete but no significant sources of
radiation have appeared in our model).

We take photoionization cross sections from \scite{verner96},
recombination rate coefficients from \scite{vf96} and \scite{arnaud85}
and collisional ionization rates from \scite{voronov97}. The cooling
rate due to collisional excitation of \HI\ was taken from
\scite{scholz91}, while that for \HeII\ was taken from \scite{black81}
with the modification introduced by \scite{cen92} at high
temperatures. The cooling rate due to free-free emission was computed
using the Gaunt factors given by \scite{sutherland98}.

\subsection{Critical Mass for Collapse}
\label{sec:mfilter}

If the IGM has a non-zero temperature, then pressure forces will
prevent gravitational collapse of the gas on small scales. In the
absence of dark matter, the effects of pressure on the growth of
density fluctuations in the gas due to their self-gravity are
described by a simple Jeans criterion, such that density fluctuations
on mass scales below the Jeans mass $M_{\rm J}$ are stable against
collapse. However, this simple criterion needs to be modified in the
case of non-linear collapse of the gas in the presence of a
gravitationally dominant cold dark matter component which collapses to
form dark matter halos. \scite{gnedin2000b} has obtained an analytical
description of the effects of gas pressure in this case, based on
earlier work by \scite{gnedin97}. Using a linear perturbation
analysis, \scite{gnedin97} found that growth of density fluctuations in
the gas is suppressed for comoving wavenumbers $k>k_{\rm F}$, where
the critical wavenumber $k_{\rm F}$ is related to the Jeans wavenumber
$k_{\rm J}$ by 
\begin{equation}
{1 \over k^2_{\rm F}(t)} = {1 \over D(t)} \int_0^t \d t^\prime
a^2(t^\prime) {\ddot{D}(t^\prime) + 2 H(t^\prime) \dot{D}(t^\prime)
\over k^2_{\rm J}(t^\prime)} \int_{t^\prime}^t {\d t^{\prime\prime}
\over a^2(t^{\prime\prime})}
\end{equation}
and $k_{\rm J}$ is defined as
\begin{equation}
k_{\rm J} =  a \left(4\pi \G \bar{\rho}_{\rm tot} {3 \mu \mh \over 5 \kb
\bar{T}_{\rm IGM}} \right)^{1/2}.
\label{eq:kF}
\end{equation}
In the above, $\bar{\rho}_{\rm tot}$ is the mean \emph{total} mass
density including dark matter, $D(t)$ and $H(t)$ are the linear growth
factor and Hubble constant respectively as functions of cosmic time
$t$, and $\dot{\mbox{}}$ represents a derivative with respect to
$t$. This expression for $k_{\rm F}$ accounts for arbitrary thermal
evolution of the IGM, through $k_{\rm J}(t)$. Corresponding to the
critical wavenumber $k_{\rm F}$ there is a critical mass $M_{\rm F}$
which we will hereafter call the filtering mass, defined as
\begin{equation}
M_{\rm F}=(4 \pi /3) \bar{\rho}_{\rm tot} (2 \pi a/k_{\rm F})^3
\label{eq:MF}
\end{equation}
The Jeans mass $M_{\rm J}$ is defined analogously in terms of $k_{\rm
J}$.  In the absence of pressure in the IGM, a halo of mass $M_{\rm
tot}$ would be expected to accrete a mass $(\Omega_{\rm
b}/\Omega_0)M_{\rm tot}$ in gas when it collapsed. \scite{gnedin2000b}
found that in cosmological gas-dynamical simulations with a
photoionized IGM, the average mass of gas $M_{\rm gas}$ which falls into
halos of mass $M_{\rm tot}$ can be fit with the formula
\begin{equation}
M_{\rm gas} =  {\left( \Omega_{\rm b} / \Omega_0 \right) M_{\rm tot} \over [1 + (2^{1/3}-1)M_{\rm F}/M_{\rm tot}]^3}
\label{eq:Mgaccr}
\end{equation}
with the same value of $M_{\rm F}$ as given by equations~(\ref{eq:kF})
and (\ref{eq:MF}). The denominator in the above expression thus gives
the factor by which the accreted gas mass is reduced because of the
IGM pressure. Specifically, $M_{\rm F}$ gives the halo mass for which
the amount of gas accreted is reduced by a factor 2 compared to the
case of no IGM pressure.

In our model, we calculate the filtering mass $M_{\rm F}(z)$ from
equations~(\ref{eq:kF}) and (\ref{eq:MF}), using for the IGM
temperature the volume-averaged value $\bar{T}_{\rm IGM}=\sum_j
f_{{\rm v},j} T_{{\rm IGM},j}$ (where $T_{{\rm IGM},j}$ is the
temperature of IGM gas in density bin $j$). We then modify the galaxy
formation model of \scite{cole2000} such that as each halo forms, it
accretes a mass of hot gas given by equation~(\ref{eq:Mgaccr}),
multiplied by a factor $(1-f_{\rm gal})$, where $f_{\rm gal}$ is the
fraction of the total baryonic mass $(\Omega_{\rm b}/\Omega_0) M_{\rm
tot}$ associated with that halo which has already formed galaxies
earlier on in progenitor halos.  The gas which would have been
accreted in the absence of IGM pressure is assumed to remain in the
IGM, but is available for accretion later on in the merging process
when another new halo is formed. The gas which does accrete is
distributed within the halo as described by \scite{cole2000}.

\subsection{Cooling Rate of Gas in Halos}
\label{sec:mappings}

The cooling of gas within dark halos, which controls how much of the
gas can collapse to form galaxies, is also affected by the ionizing
background.  The gas within dark halos is at much higher densities
than in the IGM, so we assume that it is in ionization equilibrium
under the combined effects of atomic collisions and the external
photoionizing background. We also assume that the halo gas is
optically thin to the ionizing background and to its own
emission. While in our models the mean metallicity of the IGM remains
low enough that it has negligible effect on the cooling, this is not
true for all of the gas in halos, some of which becomes significantly
metal-enriched due to ejection of gas from galaxies by supernova
feedback. We therefore use the photoionization code {\sc mappings
iii}, an updated version of the {\sc mappings ii} code used by
\scite{sutherland93}, to calculate the radiative cooling rate of gas
in halos in collisional and photoionization equlibrium, including the
effects of metals. Using this code, we calculate the net
cooling/heating rate of the gas as a function of density, temperature,
metallicity and redshift, using the photoionizing background predicted
by our model at that redshift. We also include Compton cooling due to
free electrons scattering off microwave background photons.  

\begin{figure}
\psfig{file=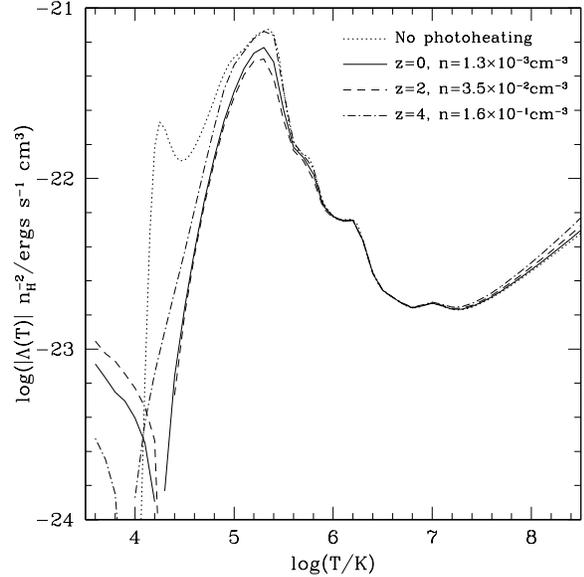,width=80mm}
\caption{The net cooling/heating function for gas at different
redshifts in the presence of the photoionizing background predicted in
our fiducial model (\S\ref{sec:fidglob}). We plot the absolute value
of the cooling - heating rate per unit volume, divided by $n_{\rm
H}^2$, for gas with metallicity $Z=0.3Z_\odot$, at redshifts $z=0$
(solid line), $z=2$ (dashed) and $z=4$ (dot-dashed). At each redshift,
we choose the gas density corresponding to the mean density in
virialized halos at that redshift (thus $n_{\rm H}=1.3\times10^{-3}$,
$3.5\times 10^{-2}$ and $1.6\times 10^{-1}$cm$^{-3}$ for $z=0$, 2, 4
respectively). The dotted line indicates the cooling curve when no
photoionizing background is present.  The $z=4$ curve is almost
indistinguishable from this case. At low temperatures ($T\approx
10^{4.3}$K) the $z=0$ and $z=2$ curves show a discontinuity, below
which there is net heating rather than cooling}
\label{fig:coolfunc}
\end{figure}

Figure~\ref{fig:coolfunc} shows the net cooling rate (i.e. the
difference of heating and cooling rates) as a function of temperature,
for gas in the presence of the ionizing background from our fiducial
model (see \S\ref{sec:fidglob}), for a metallicity $Z=0.3Z_\odot$, at
three different redshifts $z=0$, 2 and 4. The cooling rates per unit
volume are divided by $n_{\rm H}^2$, and calculated at densities
$n_{\rm H}=1.3\times10^{-3}$, $3.5\times 10^{-2}$ and $1.6\times
10^{-1}$cm$^{-3}$ at redshifts $z=0$, 2 and 4 respectively, which
correspond to the mean densities of gas in dark matter halos at those
redshifts.  We also plot the cooling curve in the absence of an
ionizing background (dotted line). For $z=0$ and $z=2$, gas cooler
than $T\approx 10^{4.3}$K is actually heated rather than cooled in the
presence of the ionizing background. We see that at $z=4$, the cooling
rate for gas at the average density for virialized halos is almost
indistinguishable from the case of zero ionizing background, a
consequence of the high gas density at this redshift, and cooling is
effective down to $T\approx 10^4$K. On the other hand, at the lower
redshifts plotted, photoionization almost completely suppresses
cooling at $T\lsim4\times10^4$K. For gas at the halo virial
temperature, the latter corresponds to a halo circular velocity of
approximately 30km/s.

In our model of galaxy formation, the gas in a dark matter halo is
assumed to be isothermal at the virial temperature $T_{\rm vir}$ of
the halo, and to have a uniform metallicity $Z_{\rm halo}$. The virial
temperature is defined in terms of the circular velocity $V_c$ at the
virial radius of the halo as
\begin{equation}
T_{\rm vir} = \frac{1}{2} \frac{\mu m_{\rm H}}{k_{\rm B}} V_c^2
\label{eq:Tvir}
\end{equation}
At each timestep in our calculations we compute the age of the halo
and the cooling time, defined as
\begin{equation}
\tau_{\rm cool} = {\frac{3}{2} n_{\rm tot} k_{\rm B} T_{\rm vir} \over
\Lambda(n_{\rm H},T_{\rm vir},Z_{\rm halo},z) }.
\label{eq:tcool}
\end{equation}
Equating $\tau_{\rm cool}$ to the age of the halo, we solve for the
density of the gas which is just able to cool, and hence for the
cooling radius, using the assumed density profile of the halo gas. We
then calculate the mass and angular momentum of gas cooling in that
timestep in the way described by
\scite{cole2000}.\footnote{Equation~4.3 of \protect\scite{cole2000}
contains a typographical error --- the factor $\mu {\rm m_H}$ should
appear in the 
numerator, not the denominator.}

\subsection{Comparison with Numerical Simulations of the IGM}

Our model of the IGM is highly simplified, but we only require it to
predict a few volume averaged quantities, namely the IGM temperature
and the spectrum of the ionizing background. The advantage of our
approach is one of speed, allowing rapid exploration of many different
models. The disadvantages, compared to N-body/gas-dynamical
simulations, are that it does not include the effects of spatial
variations in the ionizing background (no radiative transfer), and
only includes the effects of gas density variations in a very
approximate way. These limitations are likely to be most important
just prior to full reionization, when there may be large spatial variations
in the ionizing background and in the ionization state and temperature
of the IGM. However, in this paper we are interested chiefly in
calculating how an ionizing background suppresses galaxy formation,
and these suppression effects only become strong after the IGM has
been reionized, when our approximation of a uniform ionizing
background should be more accurate.

\begin{figure*}
\begin{tabular}{cc}
\psfig{file=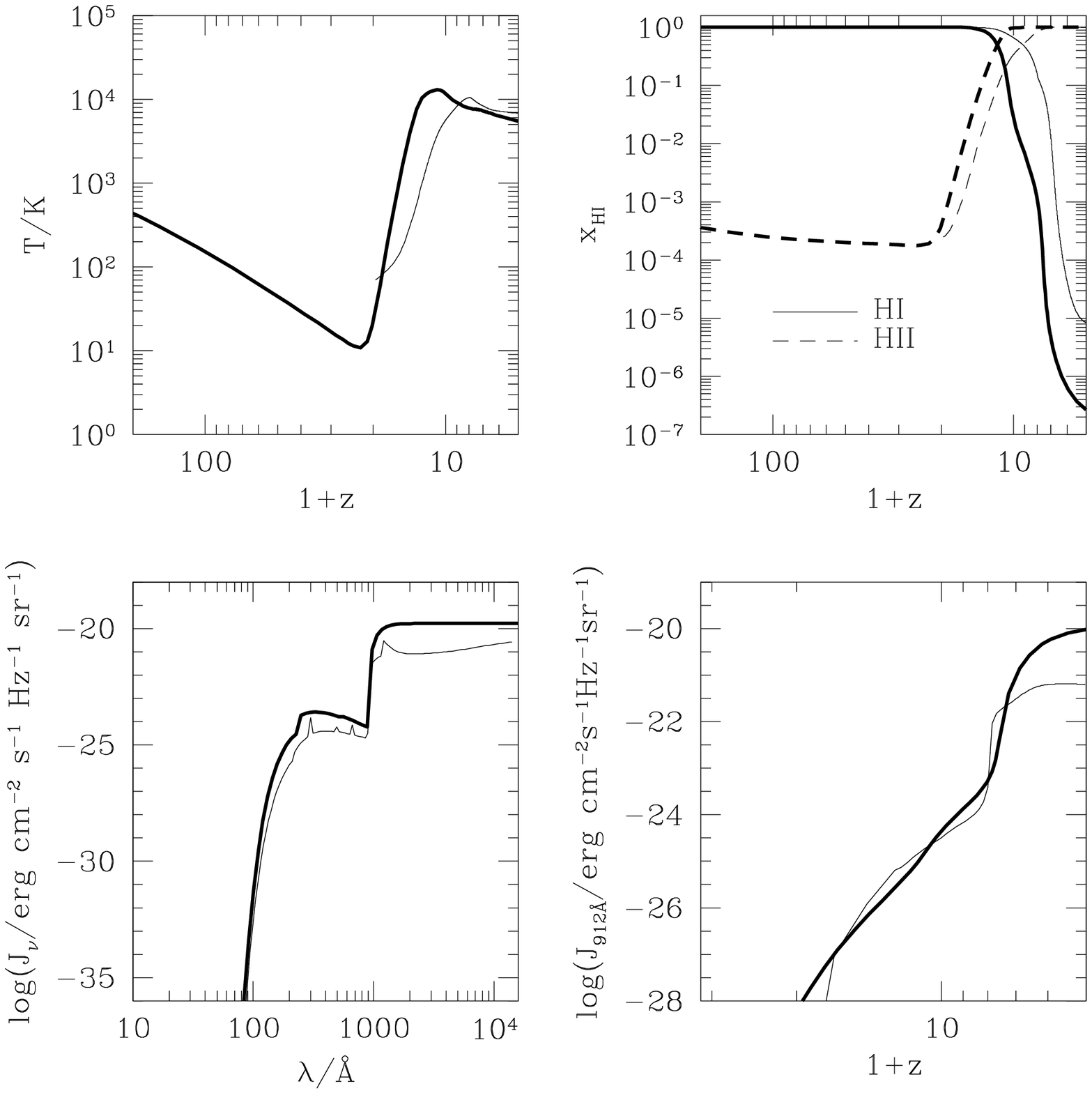,width=80mm} & \psfig{file=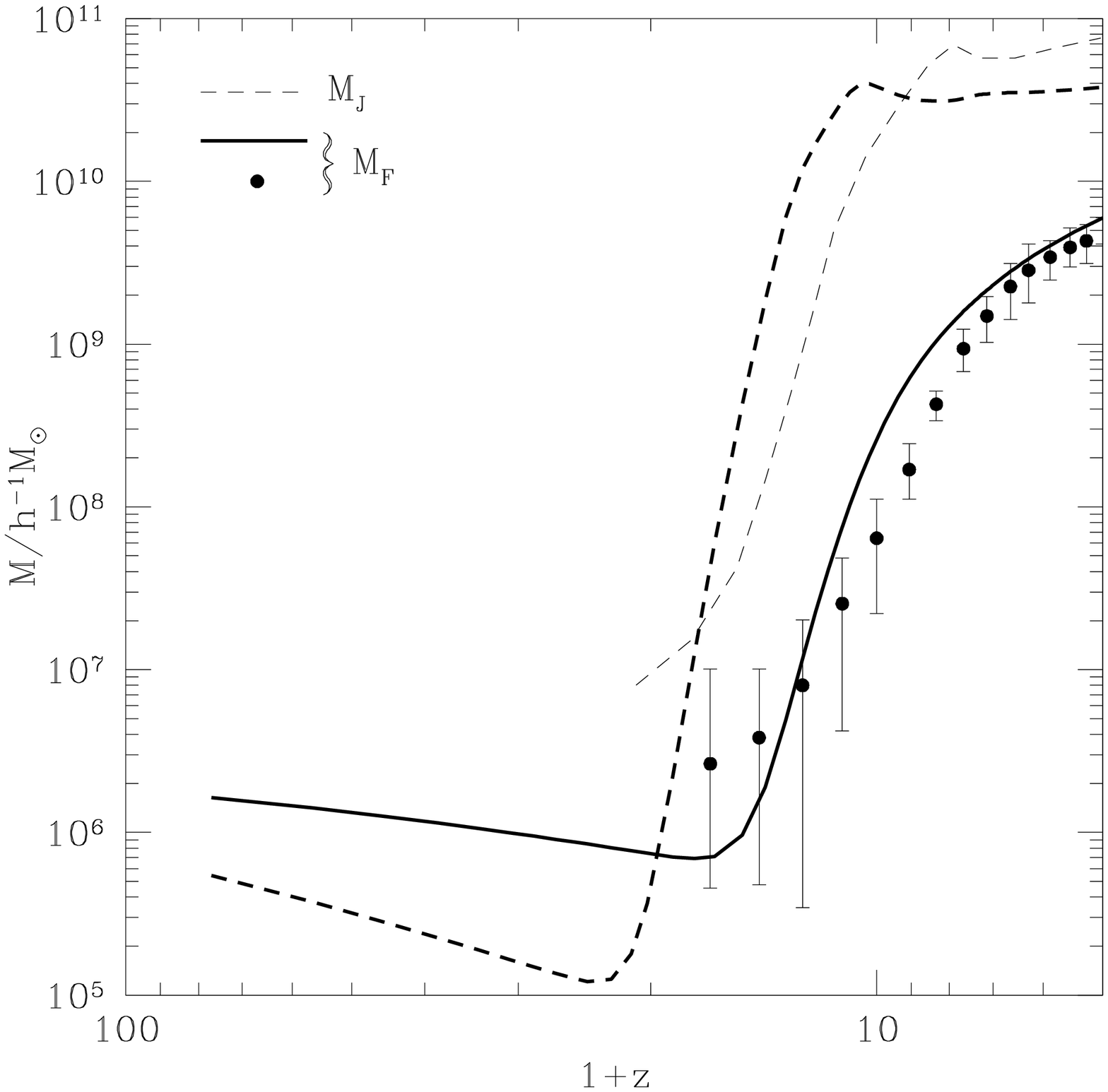,width=80mm}
\end{tabular}
\caption{A comparison of properties of the IGM and the photoionizing
background in our model (heavy lines) and in the numerical simulations
of \protect\scite{gnedin2000a} (thin lines). In our model we have
assumed the same stellar emissivity as in the simulations. In the
left-hand panel we show the volume averaged temperature (top left),
the fractions of neutral (solid lines) and ionized (dashed lines)
hydrogen (top right), the spectrum of the ionizing background at
$z\approx 9$ (bottom left) and the evolution of $J_{\nu}(912 {\rm\AA})$
with redshift (bottom right). The right hand panel compares the Jean's
and filtering masses from our model with the simulation. Dashed lines
show the Jean's mass, while points with error bars show the filtering
mass measured from the simulations and the solid line shows that
predicted by our model.}
\label{fig:gnedincompare}
\end{figure*}

We have tested the effects of the approximations in our model for the
evolution of the IGM and ionizing background by comparing it to the
N-body/gas-dynamical numerical simulations of \scite{gnedin2000a},
which include the effects of the detailed spatial distribution of gas
and ionizing sources, as well as an approximate treatment of radiative
transfer. To do this test, we input into our model the same
volume-averaged stellar emissivity and spectrum as measured from the
simulations, assuming also the same cosmological
parameters. Fig.~\ref{fig:gnedincompare} compares predictions of our
model with the same quantities measured from the simulations. Note
that Gnedin's simulation was stopped at $z=5$, so we cannot make any
comparison at lower redshift. The left-hand panel in
Fig.~\ref{fig:gnedincompare} compares the volume-averaged IGM
temperature, \HI\ and \HII\ fractions, and $J_{\nu}(912 {\rm\AA})$ as
functions of redshift, and also the background radiation spectrum at
$z=9$ (just prior to reionization for this model). The right-hand
panel compares the Jeans and filtering masses predicted by our model
with the values measured from the simulations by
\scite{gnedin2000b}. In the simulations, the filtering mass was
determined by measuring the gas masses accreted by different halos and
fitting these with the formula (\ref{eq:Mgaccr}).

Overall, the level of agreement between the two approaches is very
good, although there are some differences in detail. The temperature
of the IGM rises earlier in our model, and reionization occurs
slightly earlier, presumably because the recombination rate in
Gnedin's simulations is initially very high due to the ionizing
sources forming in the highest density regions. The ionizing
background at $z\approx 9$ is in reasonable agreement with that from
the simulation, although slightly higher. At wavelengths longwards of
912\AA\ our model predicts a significantly higher background. Here the
gas is optically thin, so the details of absorption and the
distribution of \HI\ are unimportant. It seems therefore that the
approximate radiative transfer used by \scite{gnedin2000a} somewhat
underestimates the background in the optically thin case. This is also
apparent in the bottom right panel where we show $J_{\nu}(912 \AA)$ as
a function of redshift. Prior to reionization the two models predict
very similar values, but afterwards our model reaches a significantly
higher value than does Gnedin's. In the right hand panel of
Fig.~\ref{fig:gnedincompare} we compare the Jean's and filtering
masses. The Jean's mass in our model begins to increase sooner than in
Gnedin's simulations (as expected from the earlier temperature rise in
our model), and this difference is reflected in the filtering
mass. Nevertheless, our simple model of the IGM reproduces with
reasonable accuracy the evolution of the filtering mass in the
numerical simulation. For our purposes, this is the most important
result of the comparison, because the largest effect of
photoionization on galaxy formation is through the filtering mass, as
we will see in \S\ref{sec:fidglob}.

\section{Model for the dynamical evolution of satellite galaxies}
\label{sec:tiddis}

\subsection{Model for dynamical friction and tidal stripping}

When two dark matter halos merge, a new combined dark halo is formed.
The largest of the galaxies they contained is assumed to become the
central galaxy in the new combined halo, while the other galaxies
become satellite galaxies in the new halo. These satellites evolve
under the combined effects of dynamical friction, which makes their
orbits sink towards the centre of the halo, and tidal stripping by the
gravitational field of the host halo and central galaxy, both of the
dark matter halos originally surrounding the satellites and of the
stars they contain. The \scite{cole2000} model included the effects of
dynamical friction on the evolution of satellites, but did not include
any treatment of tidal stripping. Since we are now interested in a
more detailed study of the properties of the satellites around
galaxies like the Milky Way \cite{benson01b}, we must improve our
original model to include tidal effects on satellites. We do this by
following the approach of \scite{taylor00} (with a few modifications),
following the orbits of satellites within host halos and making simple
analytical estimates of tidal effects (both ``static'' tidal
limitation and tidal ``shocks''). \scite{taylor00} show that this
simple model for the evolution of satellite halos is able to reproduce
well many of the results seen in high-resolution N-body
simulations. We describe this part of our model briefly, referring the
reader to \scite{taylor00} for a detailed discussion, but will
highlight the differences between our model and theirs.

We calculate the evolution of the orbit of each satellite galaxy in
its host halo, under the influence of dynamical friction and tidal
stripping. We specify the initial energy $E$ and angular momentum $J$
of the orbit (after the halo merger) in terms of the parameters
$R_{\rm c}^0/R_{\rm vir,host}$ and $\epsilon=J/J_{\rm C}$
respectively, where $R_{\rm c}^0(E)$ is the radius of a circular orbit
with energy $E$, $R_{\rm vir,host}$ is the virial radius of the host
halo, and $J_{\rm C}(E)$ is the angular momentum of a circular orbit
with energy $E$. We choose a constant value of $R_{\rm c}^0/R_{\rm
vir,host}$ for all satellites. Our standard choice is $R_{\rm
c}^0/R_{\rm vir,host}=0.5$, which is representative of the median
binding energy of satellite halos seen in high resolution N-body
simulations \cite{ghigna98} \emph{at the output time of the
simulation}. The typical value of $R_{\rm c}^0/R_{\rm vir,host}$ for
satellites just entering their host halo should presumably be somewhat
higher, since by the output time satellites will have lost some energy
through dynamical friction. Lacking a direct measurement of the
initial $R_{\rm c}^0/R_{\rm vir,host}$ from simulations, we will
simply use 0.5 as a default, but will also explore other values to
assess the impact of the uncertainty in this parameter on our final
results.  We select a value for the initial orbital circularity,
$\epsilon=J/J_{\rm C}$ by drawing a number at random in the range
$0.1$--$1.0$, which is a reasonable approximation to the distribution
of circularities found by \scite{ghigna98}. These choices for the
initial orbital energy and angular momentum are the same as those of
\scite{bullock00}. Given the energy and angular momentum of the orbit,
we determine the apocentric distance and begin integration of the
orbit equations at that point, where tidal forces are weakest.

We model the dark matter in both the host and satellite halos as an
NFW density profile \cite{NFW}, modified by the gravity of the galaxy
which has condensed at the halo centre (the calculation of this
adiabatic compression of the halo is described in detail by
\pcite{cole2000}).  The galaxy at the centre of each halo is modelled
as a combination of disk and spheroid. The disk has a density
distribution given by
\begin{equation}
\rho_{\rm d}(x,y,z)=\rho_{\rm d,0} \exp\left[
-\frac{ (x^2+y^2)^{1/2} }{ r_{\rm d} } \right] 
{\rm sech}^2 \left({z \over h r_{\rm d}}\right), 
\label{eq:diskdens}
\end{equation}
where $r_{\rm d}$ is the disk radial scale length, and $h$ is the
ratio of vertical to radial scale-length, which we take to be constant
and equal to 0.1. The spheroid is modelled as a spherically symmetric
$r^{1/4}$-law. The masses and sizes of these components are determined
as described by \scite{cole2000}.

The satellite galaxy+halo moves under the influence of two forces. The
first is just the net gravitational force due to the host halo and its
central galaxy. The force due to the disk is calculated using the
method of \scite{kuijken89}. The second force is that due to dynamical
friction, which we estimate using Chandrasekhar's formula
(e.g. \pcite{bt87}, section~7.1)
\begin{equation}
{\bf F}_{{\rm df},i}=-4\pi \G^2 M_{\rm s}^2 \ln \Lambda_i \rho_i B(x) {{\bf v}_{{\rm rel},i} \over |{\bf v}_{{\rm rel},i}|^3 },
\end{equation}
where $\ln \Lambda_i$ is the Coulomb logarithm, $\rho_i$ the local
density, $B(x)={\rm erf}(x)-2x\exp(-x^2)/\sqrt{\pi}$, $x=|{\bf v}_{\rm
rel}|/\sqrt{2}\sigma_i$, $\sigma_i$ is the velocity dispersion and
${\bf v}_{{\rm rel},i}$ is the relative velocity of the satellite and
component $i$. We consider two components which contribute dynamical
friction forces, namely the dark matter of the host halo and the
spheroid of the central galaxy (which we treat together and indicate
hereafter by a subscript ``h''), and the disk of the host halo galaxy
(indicated by a subscript ``d''). We adopt the same values for the
Coulomb logarithms as did \scite{taylor00} (namely 2.4 for the dark
matter/spheroid and 0.5 for the disk), which fit the results of N-body
simulations well. \scite{taylor00} discuss in detail the possible
choices for the Coulomb logarithms. While previous semi-analytic
models have often used $\ln \Lambda_{\rm h}=\ln M_{\rm h}/M{\rm s}$,
we prefer to use the same value as \scite{taylor00} for this present
work. The dynamical friction force depends upon the mass $M_s$ of the
satellite. We include in this mass that part of the satellite galaxy
and its dark halo which has not yet been stripped by tidal forces.

For
the disk velocity dispersion, we take $\sigma_{\rm d}=V_{\rm
c}/\sqrt{2}$, where $V_{\rm c}$ is the rotation speed of the disk
(computed for a spherically averaged disk), as did
\scite{taylor00}. This results in an unrealistically high velocity
dispersion when applied to the Milky Way (where the observed 1-D
velocity dispersion is approximately 30 to 40 ${\rm kms^{-1}}$). We
prefer to use 
the \scite{taylor00} value at present, but find that using a lower
value ($\sigma_{\rm d}=0.2 V_{\rm c}$) has almost no effect on the
results presented in this paper (e.g. the galaxy luminosity functions
of Fig.~\ref{fig:fidLF} are hardly affected by this change). For the
dark matter/spheroid system we find $\sigma_{\rm h}$ by integration of
the Jean's equation (assuming an isotropic velocity dispersion)
\begin{equation}
{\d (\rho_{\rm h}\sigma_{\rm h}^2) \over \d r} = -{\G M_{\rm h}(r) \over r^2}
\rho_{\rm h}(r),
\label{eq:jeans}
\end{equation}
where $M_{\rm h}$ is the total (i.e. dark plus baryonic) mass within
radius $r$ of the host halo. We assume that that dark matter follows
the NFW profile for all radii outside of the virial
radius. \scite{cl96} show that the velocity dispersion calculated in
this way is in reasonable agreement with that measured in N-body
simulations.

At each point in the orbit, we calculate the ``static'' tidal
limitation radius of the satellite galaxy+halo, $r_{\rm t}$. This is
the radius where the gravitational force of the satellite equals the
sum of the tidal force from the host halo plus the pseudo-force due to
the satellite's orbit,
\begin{equation}
{\G M_{\rm s}(r_{\rm t}) \over r_{\rm t}^2}=\left[ \omega^2 - {\d \over \d R} \left( {\G M_{\rm h}(R)\over R^2} \right) \right] r_{\rm t},
\label{eq:tidforce}
\end{equation}
where $R$ is the distance from the centre of the host halo, $M_{\rm
s}(r_{\rm t})$ is the total mass within radius $r_{\rm t}$ of the
satellite, and $\omega$ is the instantaneous angular velocity of the
satellite. Note that the factor of $\omega^2$ is strictly accurate
only for circular orbits. Here we follow \scite{taylor00} and include
this term for all orbits. For the purposes of this calculation and
that of $\sigma_{\rm h}$, the mass of the host halo disk is
spherically averaged (the assumption under which
eqn.~\ref{eq:tidforce} was derived). In all other calculations of
satellite dynamics we use the density distribution of
eqn.~(\ref{eq:diskdens}) to describe the
disk. Equation~(\ref{eq:tidforce}) is valid under the assumption that
the satellite is much smaller than the host halo, which is true for
all but a very small fraction of satellites in our
calculations. 

\scite{weinberg94} has argued that mass loss may occur at smaller
radii than suggested by the above expression, due to heating of the
satellite by gravitational shocking as it passes near the centre of
the host halo. We adopt the approach of \scite{taylor00} to estimate
the effect of this tidal shocking, and refer the reader to that paper
for a complete description of the method. Briefly, during each fast
shock (i.e. any shock for which the timescale is less than the
internal orbital period of the satellite at its half-mass radius), we
calculate the rate of heating by tidal forces. The energy thereby
deposited in the satellite causes the satellite to expand, pushing
some material beyond the tidal radius, and so allowing more material
to be removed by tidal forces. We define the {\em effective} tidal
radius $r_{\rm t}$ as the radius in the original satellite density
profile beyond which material has been lost. When the satellite has
been heated, this effective tidal radius will therefore be less than
$r_{\rm t}$ as defined by eqn.~(\ref{eq:tidforce}). We remove matter
from the satellite in spherical shells outside of the effective tidal
radius in the heated satellite. Note that we leave the density profile
of material inside the effective tidal radius unchanged, so that the
maximum circular velocity in the satellite remains unchanged until the
effective tidal radius is reduced below the position of the peak of
the rotation curve (i.e. $2.16r_{\rm s}$ for a pure NFW dark matter
rotation curve, but some other value when the baryonic contribution is
included).

Mass beyond the effective tidal radius of the satellite is removed
gradually on the shorter of the angular orbital timescale, $2
\pi/\omega$ (which becomes the orbital period for circular orbits),
and the radial infall timescale $R/v_{\rm R}$.  \scite{taylor00} only
considered the angular orbital timescale, which results in very low
mass loss rates for satellites on nearly radial orbits (as sometimes
occur if the dynamical friction force is strong). It is then simple to
calculate the mass remaining in the satellite (including both dark
matter and baryonic components) and to use this to calculate the
dynamical friction force exerted on the satellite.

The orbit equations are integrated until one of three conditions is
met:
\begin{enumerate}
\item The final redshift (i.e. the redshift at which we are studying
the galaxy population) is reached. In this case we calculate the
remaining mass and luminosity of the satellite galaxy after tidal
limitation.
\item The host halo merges to become part of a new halo. The satellite
halo then becomes a satellite of the new halo and is assigned a new
orbit in that halo. We begin integration of the orbit equations again,
but starting with the previous value of the effective $r_{\rm t}$ for
the satellite.
\item The satellite merges with the central galaxy (which we assume
happens when the orbital radius, $R$, first reaches $R_{\rm merge}$,
which we take to be the sum of the half-mass radii of the host and
satellite galaxies\footnote{The model results are insensitive to the
exact definition of merger time, as once $R$ reaches such small radii
it decreases very rapidly to zero.}). In this case we add to the
central galaxy of the host halo the remaining mass of satellite galaxy
at the time when the pericentre of its orbit first passed within
$R_{\rm merge}$ (even though some of this mass may have been stripped
off since that time, its orbit will carry it into the central galaxy in
any case), using the rules described by \scite{cole2000} (and possibly
triggering a burst of star formation).
\end{enumerate}

Stellar mass stripped from satellite galaxies is added to a diffuse
stellar component of the host halo, but is not considered further in
our models. Any cold gas stripped from the satellites is added to the
hot gas reservoir of the host halo, and so may be able to cool again
at a later time. Finally, as the satellite galaxy orbits in the host
halo it continues to form stars, which causes some of the cold gas
mass of the galaxy to be ejected according to the supernova feedback
prescription of \scite{cole2000}. In the case of satellites with
shallow potential wells, this can significantly alter the mass of the
galaxy along its orbit. Therefore the mass of this reheated gas
is removed from the satellite halo during the orbit.

\begin{figure}
\psfig{file=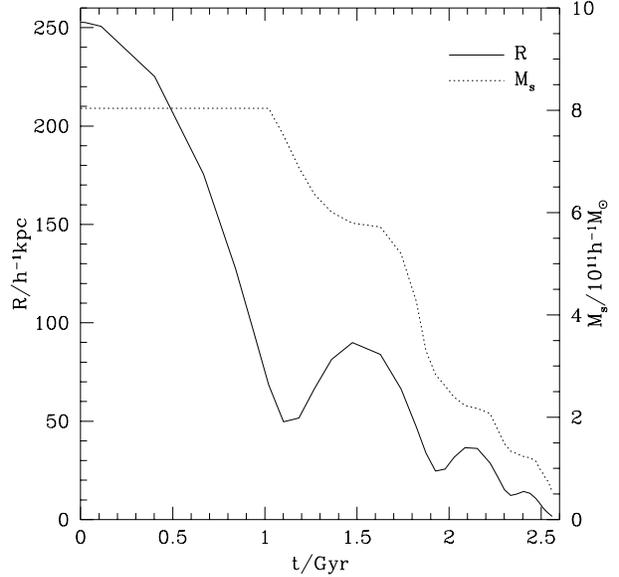,width=80mm}
\caption{An example of the evolution a satellite galaxy orbit. The
satellite enters the host halo at $t=0$ and merges with the central
galaxy of that halo after approximately 2.5~Gyr. The solid line shows
the orbital radius of the satellite as a function of time, showing
that the orbit is decaying rapidly due to the effects of dynamical
friction. The dotted line shows the remaining mass of the satellite.
Note that the mass does not begin to decrease until the first passage
through pericentre, as before this the tidal forces felt by the
satellite are not strong enough to remove any mass. See text for more
details.}
\label{fig:orbit}
\end{figure}

In Fig.~\ref{fig:orbit} we show an example of a satellite orbit
calculated using the above model. The host halo has a mass of $2\times
10^{13}h^{-1}M_\odot$, a virial circular velocity of 440km/s and
concentration (defined here as the ratio of virial radius to NFW scale
radius) of 5.9. The same three quantities for the satellite when it
was still a separate halo are $3.5\times 10^{12}h^{-1}M_\odot$,
340km/s and 5.6, respectively. However, in this example, the satellite
halo has already lost mass while being a satellite in a progenitor of the
current host halo, which is why in this plot it starts from a mass of
$8\times 10^{11}h^{-1}M_\odot$. In this plot, the time $t$ is measured
from when the host halo formed, and the satellite orbit begins at the
apocentre (approximately $250h^{-1}$kpc from the centre of the host
halo). The orbit decays rapidly due to the effects of dynamical
friction, so that the satellite makes three orbits before merging with
the galaxy at the centre of the host halo at $t\approx 2.5$Gyr. The
mass of the satellite is seen to decrease most rapidly when the
satellite is close to pericentre. Note also that until just before the
first passage through pericentre, the mass of the satellite is
unchanging, as before this time tidal forces are simply not strong
enough to strip any mass from the halo.

We do not attempt here to model changes in the density profile of the
satellite galaxy+halo within the tidal radius -- the profile of the
unstripped material is assumed to remain as it was before any
stripping occurred. Nor do we account for any changes in the
global properties of a galaxy which has lost mass to tidal forces
(i.e. the galaxy keeps the same scale lengths, star formation
timescale etc. as it had before any mass loss occurred). Numerical
simulations of satellites undergoing tidal interactions
(e.g. \pcite{mayer01}) demonstrate that satellite mass profiles are
affected by tidal interactions. Typically, they find a reduction in
the amount of stellar mass within a given radius for radii within but
comparable to the tidal radius. For low surface brightness galaxies
(LSBs), this reduction can be up to a factor of around 2, but the
effect is much weaker for high surface brightness galaxies (HSBs). At
small radii, the stellar mass within a fixed radius is often increased
by tidal interactions (through the production of a bar). The rotation
curves of the galaxies are more seriously affected (presumably because
the spherically distributed dark matter is less strongly bound than
the stellar disk), often being reduced by a factor 2 for LSB galaxies
(and somewhat less for HSB galaxies).

Our calculation of merging times improves upon the simple estimates
previously used in many semi-analytic models (which have often used
results for satellites orbiting in isothermal halos, with no tidal
stripping). However, we find that on average our approach predicts
comparable merging timescales for satellite halos to the simpler
treatment in \scite{cole2000}, although some fraction of satellites
are predicted to have extremely long timescales, as they lose so much
mass through tidal stripping that dynamical friction forces become
extremely weak.

\begin{figure*}
\psfig{file=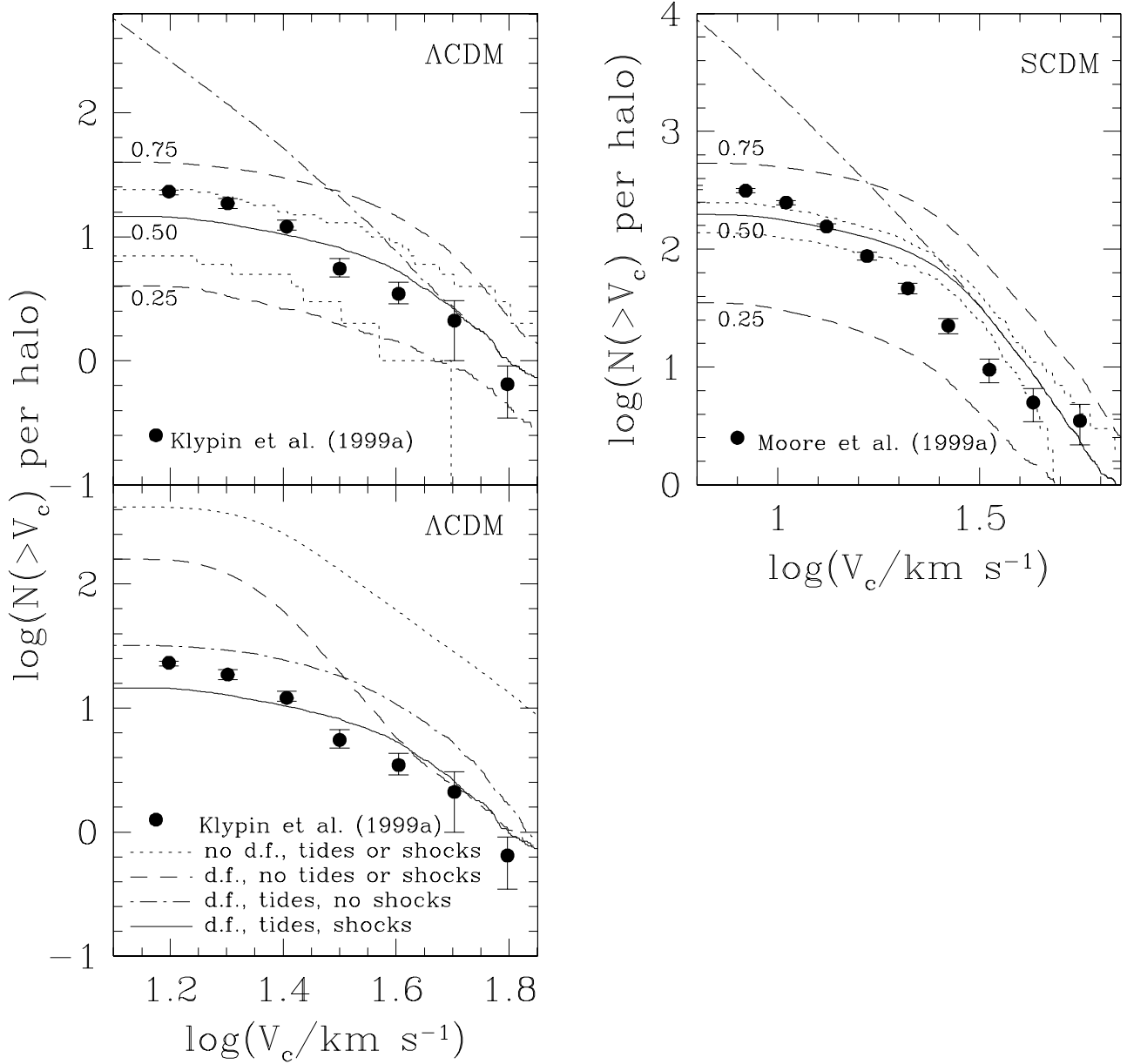,width=180mm,bbllx=0mm,bblly=90mm,bburx=188mm,bbury=270mm,clip=}
\caption{The number of satellite halos as a function of circular
velocity from the semi-analytic model compared to N-body simulations,
for Milky Way-like halos in CDM models. $N(>V_c)$ is the cumulative
number of sub-halos per host halo, with $V_c$ defined as the peak
circular velocity of the sub-halo. The simulations are of a
$\Lambda$CDM model from \protect\scite{klypin99a} and of a SCDM model
from \protect\scite{moore99}. In each panel, the solid points with
error bars show the N-body simulation results, while the lines show
the semi-analytic predictions for different assumptions. {\em Upper
two panels:} the light solid lines show the semi-analytic prediction
(averaged over 300 realizations), including in
the semi-analytic model the same cut on sub-halo mass as in the
simulations. The dotted lines on either side of the solid line show
the 10\%-90\% range of the distribution around the mean value. The
dot-dashed lines show the result from the semi-analytic model when no
mass cut is applied. These results are for $R_{\rm c}^0/R_{\rm
vir,host}=0.5$. The dashed lines show the results if instead we assume
$R_{\rm c}^0/R_{\rm vir,host}=0.75$ or $0.25$, including tidal
stripping and the mass cut. {\em Lower panel:} this shows the
contribution of different physical processes in the semi-analytic
model, for the case of $\Lambda$CDM. The dotted line shows the
predicted sub-halo velocity distribution for the case of no dynamical
friction, no static tidal limitation and no tidal shocking. Switching
on dynamical friction produces the dashed line. Adding in static
tidal limitation gives the dot-dashed line, and finally switching on
tidal shocking produces the solid line. In this panel, all the curves
are for $R_{\rm c}^0/R_{\rm vir,host}=0.5$, and all include the same
cut on sub-halo mass as in the N-body simulation.}
\label{fig:satvs}
\end{figure*}

\subsection{Comparison with N-body simulations}

For our present purposes, we are most interested in whether our model
reproduces the abundance of satellite halos, (or sub-halos) in a host
halo typical of Milky Way-like galaxies. We therefore compare the
predictions of our model for the number of satellite halos with the
results from high resolution, dark-matter-only N-body simulations of
Milky Way-like halos in CDM models. Fig.~\ref{fig:satvs} shows the
comparisons with simulations of $\Lambda$CDM ($\Omega=0.3$,
$\Lambda=0.7$) by \scite{klypin99a} and of SCDM ($\Omega=1$) by
\scite{moore99}. In the $\Lambda$CDM simulation, the host halo mass is
$M_{\rm halo}=1.1\times 10^{12}$, and the minimum resolvable sub-halo
mass is $3.3\times 10^8h^{-1}M_\odot$ (corresponding to 20
particles). The corresponding quantities in the SCDM simulation are
$1.0\times 10^{12}h^{-1}M_\odot$ and $1.6\times 10^7h^{-1}M_\odot$. We
make the comparison in terms of the sub-halo velocity function
$N(>V_c)$, defined as the cumulative number of sub-halos per host halo
with circular velocities greater than $V_c$, where $V_c$ is defined as
the peak circular velocity of the sub-halo.

To compare our results to those of the N-body simulations, we run our
semi-analytic model without baryons. Each dark matter halo then has a
pure NFW profile. In the case of satellite halos, the NFW profile is
truncated beyond a radius $r_t$ determined by the combined effects of
static tidal limitation and tidal shocking. For an untruncated NFW
profile, the circular velocity peaks at $r_m=2.16 r_s$, where $r_s$ is
the NFW scale radius. We assume that the density profile of satellite
halos is unchanged within $r_t$, so the peak $V_c$ is $V_{\rm
NFW}(r_m)$ if $r_t>r_m$, and $V_{\rm NFW}(r_t)$ otherwise. We choose a
host halo mass at $z=0$ equal to the value in the simulation, run 300
different realizations of the halo merger tree, and then take the mean
$N(>V_c)$ averaged over these realizations. In our model, sub-halos
are only completely destroyed when they merge into the centre of the
host halo. Tidal stripping reduces the mass of a halo, but is assumed
never to destroy it completely. In our semi-analytic model, we can
resolve much lower mass halos than can be resolved in the N-body
simulations. Since there can be a wide range of sub-halo masses at a
given value of the sub-halo circular velocity $V_c$, it is essential
to take into account this difference in mass resolution when we
compare to the simulations. Therefore, to calculate $N(>V_c)$, we
discard from the semi-analytic model all sub-halos with masses (within
$r_t$) smaller than the minimum resolvable sub-halo mass in the
simulation. As Fig.~\ref{fig:satvs} shows (compare the solid and
dot-dashed curves), this mass cut produces a large reduction in the
number of satellites below $V_c = 20-40\, {\rm km s^{-1}}$, and is very
important for matching the simulation results.

The upper left and right panels of Fig.~\ref{fig:satvs} show the
comparison of our model with the $\Lambda$CDM and SCDM simulations
respectively. The solid curves show the prediction for $R_{\rm
c}^0/R_{\rm vir,host}=0.5$, including the mass resolution cut. The
dotted lines on either side of the solid line show the 10\%-90\% range
of the distribution seen among the different realizations. This range
is larger for the $\Lambda$CDM than SCDM model, which mainly results
from the smaller number of sub-halos per host halo in the former
case. In the same panels, the upper and lower dashed lines show the
effect of changing the assumed initial orbital energy to $R_{\rm
c}^0/R_{\rm vir,host}=0.75$ and $0.25$ respectively. We see that our
standard value $R_{\rm c}^0/R_{\rm vir,host}=0.5$ gives significantly
better agreement with the N-body simulations for both $\Lambda$CDM and
SCDM.

The lower left-hand panel of Fig.~\ref{fig:satvs} shows in more detail
the separate effects of dynamical friction, static tidal limitation
(i.e. $r_{\rm t}$ as defined by eqn.~\ref{eq:tidforce}) and tidal
shocks on the velocity distribution, for the $\Lambda$CDM case. All of
the curves plotted there assume $R_{\rm c}^0/R_{\rm vir,host}=0.5$,
and include the cut in sub-halo mass corresponding to the resolution
of the N-body simulation.  The solid line includes all of the above
processes, and is therefore identical to the solid line in the upper
left-hand panel. The dotted line contains none of these processes, so
sub-halos never merge and are never tidally stripped. Switching on
dynamical friction results in the dashed line, which greatly reduces
the number of high $V_{\rm c}$ (relatively massive) subhalos, but is
much less important for the low $V_{\rm c}$ halos. Switching on static
tidal limitation (and keeping dynamical friction switched on) results
in the dot-dashed line. This greatly reduces the number of low $V_{\rm
c}$ halos, as these are strongly affected by tidal forces, once the
mass cut is included. The number of high $V_{\rm c}$ halos actually
increases somewhat, since tidal limitation is able to reduce the mass
of these halos and so reduce the strength of the dynamical friction
forces which they experience. The remaining difference between the
dot-dashed and solid lines is accounted for by tidal shocking. The
overall effect of tidal stripping is to reduce the number of halos at
low $V_c$ by a factor $\sim 10$.

An important difference between our model and the similar calculation
by \scite{bullock00} is we can resolve sub-halos within sub-halos
(i.e. we record all branches of the merger tree and so halos merging
with the final halo may have substructure of their own), whereas
Bullock et al. considered only sub-halos (i.e. merging halos were
assumed to have no substructure of their own). This distinction is
important, since, in our model of galaxy formation, every branch of the
merger tree can potentially host a galaxy (providing its virial
temperature exceeds $10^4$K and so is able to cool
efficiently). Hence, when no tidal stripping is applied we find many
more satellites at a given $V_{\rm c}$ than did Bullock et
al. However, when tidal stripping is included, both our model and that
of Bullock et al.  are in reasonable agreement with the N-body results
(although Bullock et al. compared only with the $\Lambda$CDM
simulations of \scite{klypin99a}).

We remind the reader that our choice of $R_{\rm c}^0/R_{\rm vir}=0.5$
was originally motivated by the measurement of the orbital energy
distribution of all the satellites existing in a halo at the final
output time of an N-body simulation. However, in our model, we use
this $R_{\rm c}^0/R_{\rm vir}$ as the initial value for each satellite
after it joins the main halo. Plausibly we should use a higher value,
since when satellites first fall into a host halo they should be less
bound than at any subsequent time. This would reduce the effectiveness
of tidal limitation in our model (e.g. compare the curves for $R_{\rm
c}^0/R_{\rm vir}=0.50$ and 0.75 in Fig.~\ref{fig:satvs}). Also, as
noted above, we do not include any adjustment in the density profile
of the material within the tidal radius in response to stripping of
material from larger radii. This would be expected to make satellites
less bound and to enhance the process of tidal stripping, and also to
lower the circular velocity. These two effects work in opposite
directions, but it is not clear which is the dominant process. N-body
suggestions suggest that the effect of the latter on the sub-halo peak
circular velocities $V_c$ is in fact fairly small; \scite{ghigna00}
find in their high-resolution simulations that for sub-halos where the
tidal radius is larger than the initial peak-$V_c$ radius, $V_c$
typically changes by only $\sim 20\%$ due to tidal effects. For now,
we simply note that $R_{\rm c}^0/R_{\rm vir}=0.50$ does produce
reasonable agreement with the numerical results, and so we adopt this
throughout the remainder of this paper.

Our semi-analytic model includes the effects of baryonic collapse on
the mass profiles of the host and satellite halos, although this
effect is turned off when we compare to pure dark matter N-body
simulations.  While baryonic dissipation makes satellite halos more
strongly bound, and so more resistant to tidal limitation, it also
makes the tidal forces of the host halo and central galaxy
stronger. The central galaxy of the host halo also contributes to
satellite destruction through its contribution to dynamical
friction. In halos of mass $\sim 10^{12}h^{-1}M_\odot$, the net result
of including the baryonic components is to further reduce the number
of satellite halos (compared to a pure dark matter calculation). We
find, for these halos and with our standard galaxy formation model
(see \S\ref{sec:fidglob}), that the number of satellites at a given
$V_{\rm c}$ is reduced by around 40\% at $V_{\rm c}=60$km/s, and by
about 60\% at $V_{\rm c}=15$km/s.

\section{Results}
\label{sec:results}

We are now able to explore in a self-consistent way the effects of a
photoionizing background on the properties of galaxies and also the
effects of galaxies on the IGM. We will begin in \S\ref{sec:IGMprops}
by obtaining a self-consistent model and exploring the evolution of
the IGM and ionizing background. In \S\ref{sec:fidglob} we examine the
effects on the global properties of galaxies in the fiducial model of
\scite{cole2000}. The properties of satellite galaxies will be
explored in a separate paper \cite{benson01b}.

\subsection{Properties of the IGM and Ionizing Background}
\label{sec:IGMprops}

\subsubsection{Star Formation History}
\label{sec:sfrhist}

Our starting point is the fiducial model of \scite{cole2000}, modified
in the ways described in \S\ref{sec:model} and \S\ref{sec:tiddis}. We use
this to predict the star formation history and associated emissivity
in ionizing photons as a function of redshift. We resolve all halos
that are able to cool in the redshift interval 0 to 25 to ensure that
all ionizing photons are accounted for. To determine the spectrum of
emission from these stars, we tabulate the mean star formation rate
per unit volume from our model as a function of both cosmic time and
metallicity, $\d^2 \rho_\star(t,Z)/\d t\, \d Z$. The stellar emissivity per
unit volume at cosmic time $t$ is then simply
\begin{equation}
F_{\lambda}(t)=\int_0^t \int_0^\infty {\d^2
\rho_\star(t^\prime,Z)\over\d t^\prime \d Z} \Phi(t-t^\prime,Z)\, \d
Z\, \d t^\prime ,
\label{eq:fstellar}
\end{equation}
where $\Phi(t,Z)$ is the spectral energy distribution of a stellar
population of age $t$ and metallicity $Z$, which we take from the
models of \scite{bc99}. 

To account for the effects of absorption by dust and gas in galaxies
on the ionizing emissivity, we multiply the above expression
(\ref{eq:fstellar}) at wavelengths $\lambda < 912 {\rm\AA}$ by a constant
factor $f_{\rm esc}$, defined as the fraction of ionizing photons
produced by stars that escape through the dust and gas of the galaxy's
interstellar medium (ISM). To calculate the effects of dust absorption
for the non-ionizing radiation at $\lambda > 912{\rm\AA}$, we use the same
approach as in \scite{cole2000}.  The value of $f_{\rm esc}$ for the
ionizing photons is uncertain, so we will present results for two
values, $f_{\rm esc}=10\%$ and $f_{\rm esc}=100\%$, which we believe
to bracket a reasonable range. The value $f_{\rm esc}=100\%$ results
in our model in an emissivity in Lyc ($\lambda < 912{\rm\AA}$) photons from
galaxies at $z=3$ that agrees with the recent observational estimate
for Lyman-break galaxies by \scite{steidel00} (after allowing for the
differences in the assumed cosmological models), and predicts
reionization of hydrogen at $z\approx 8$, compatable with measurements
of the Gunn-Peterson effect in quasars. On the other hand, the value
$f_{\rm esc}=10\%$ is more consistent with both observational
(e.g. \pcite{leitherer95,steidel00}) and theoretical
(e.g. \pcite{dsf00,benson00c}) estimates of the escape fraction at
both low and high redshift. The observational estimate of
$L_{\nu}(900{\rm\AA})/L_{\nu}(1500{\rm\AA})$ for the Lyman-break galaxies by
\scite{steidel00} implies $f_{\rm esc}\sim 10-40\%$, allowing for
uncertainties in dust extinction and in the emission at $\lambda <
912\AA$ predicted by stellar models. The reason why we require a
larger $f_{\rm esc}$ than \scite{steidel00} to produce the same net
ionizing emissivity at $z=3$ is that the \scite{cole2000} model
predicts too low a typical 1500\AA\ luminosity for Lyman-break
galaxies, once we include dust. Our model with $f_{\rm esc}=10\%$
predicts an ionizing background that is in better agreement with
observational estimates at $z<4.5$, but also predicts a low redshift
for reionization, $z\approx 5$, that is barely compatable with the
Gunn-Peterson constraints. These issues are discussed in more detail
below. Since it is the redshift of hydrogen reionization and reheating
of the IGM that appears to be the most important in determining the
effects on galaxy formation, we will take $f_{\rm esc}=100\%$ as our
standard case.

We also include the contribution to the ionizing emissivity from quasars,
according to the observational parameterization of
\scite{madau99}. Their parameterization is based on fitting
observational data on numbers, magnitudes and redshifts of quasars at
$z<4.5$, assuming an Einstein-de Sitter cosmology. To obtain the
emissivity in our chosen cosmology, we must allow for the dependence
of the observationally-inferred luminosities and number densities on
the assumed cosmological model. We therefore use
\begin{equation}
\epsilon(z) = \epsilon_{\rm MHR}(z) \left({d_{\rm L}(z) \over d^{\rm
(EdS)}_{\rm L}(z)} \right)^2 \left( {\d V(z)/\d z^{\rm (EdS)} \over \d
V(z)/\d z} \right), 
\end{equation}
where $\epsilon_{\rm MHR}(z)$ is the emissivity from \scite{madau99}
(measured from their Fig.2), $d_{\rm L}(z)$ is the luminosity distance
and $\d V(z)/\d z$ is the comoving volume per unit redshift. Functions
with superscript (EdS) are calculated in the Einstein-de Sitter
cosmology, those without in the cosmology of our fiducial model. We
use the same expression to extrapolate the quasar contribution to
$z>4.5$. We note that even at $z<4.5$, the use of $\epsilon_{\rm
MHR}(z)$ from \scite{madau99} involves a considerable extrapolation of
the quasar luminosity function down to luminosities not directly
observed. 

We then use the total (i.e. stellar plus quasar) ionizing emissivity
in calculating the thermal evolution of the IGM (and hence the
filtering mass) and the ionizing background, both as functions of
redshift. The stellar emissivity must be determined self-consistently
with the feedback effects on galaxy formation from the IGM pressure
and ionizing background, as described in \S\ref{sec:mfilter} and
\S\ref{sec:mappings}. We do this by means of an iterative procedure,
starting from a galaxy formation model computed ignoring these
feedback effects, calculating the ionizing background in this model,
using this as input in calculating a revised model including the
photoionization feedback effects, and repeating this cycle until we
have a model whose star formation history is consistent with the
photoionizing background that it produces. \scite{cole2000} chose the
parameters of their fiducial model to match certain observations of
the local galaxy population, in particular the luminosity function of
galaxies in the B and K-bands. We find that if we keep the same
parameter values as used by \scite{cole2000}, then when we include the
photoionization feedback, our model still produces an acceptable fit
to these luminosity functions.  The only change is a small adjustment
of $\Upsilon$ (which determines mass-to-light ratios) from 1.38 in
\scite{cole2000} to 1.32 (we adjust the recycled fraction in our
chemical evolution model accordingly).  As will be discussed in more
detail in \S\ref{sec:fidLF}, the faint-end slopes are now somewhat
flatter than before, but this is consistent with recent determinations
of the luminosity functions.

\begin{figure}
\psfig{file=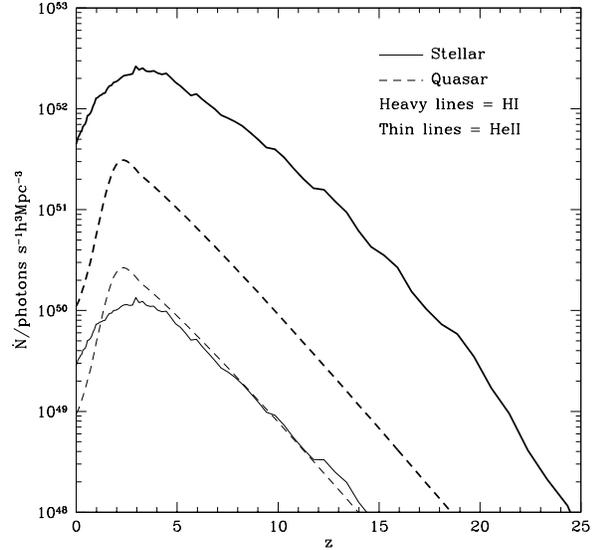,width=80mm}
\caption{The emissivities in \HI\ and \HeII\ ionizing photons (heavy
and thin lines respectively) per comoving volume as a function of
redshift. Solid lines show the emissivity from stars assuming $f_{\rm
esc}=100\%$, while dashed lines show that from quasars.}
\label{fig:emissivity}
\end{figure}

The net emissivity in \HI\ and \HeII\ ionizing photons from stars and
quasars for our standard model is shown in
Fig.~\ref{fig:emissivity}. For $f_{\rm esc}=100\%$ stars always
dominate the production of \HI\ ionizing photons, but quasars dominate
the production of \HeII\ ionizing photons until $z\approx 1$ when the
rapidly falling quasar emissivity leads to stars becoming the dominant
source.

\begin{figure}
\psfig{file=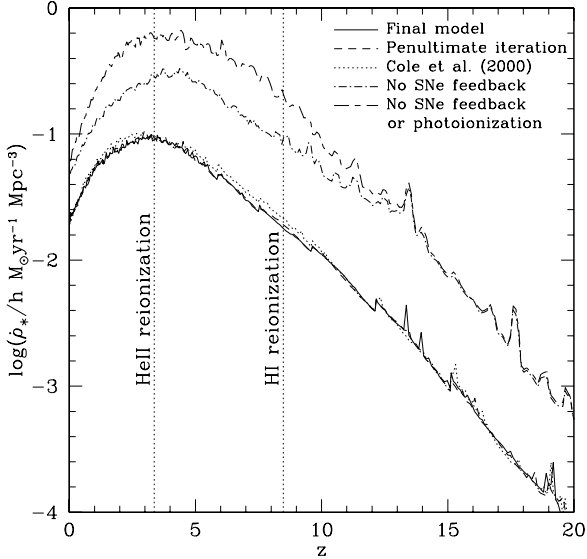,width=80mm}
\caption{The star formation rate per comoving volume in our standard
model as a function of redshift. The solid line represents the star
formation rate used to compute the temperature of the IGM and the
evolution of the ionizing background. The dashed line indicates the
star formation rate in the penultimate iteration of the model,
indicating that convergence has been reached over the range of
interest. The dotted curve indicates the star formation rate in the
model of \protect\scite{cole2000} scaled to the value of $\Upsilon$ in
our standard model for the purposes of this comparison. We also show a
model with the effects of supernovae feedback switched off, both with
(dot-dashed line) and without (short-dashed-long-dashed line) the effects
of photoionization feedback included. The epochs of \HI\ and \HeII\
reionization in the standard model are marked by vertical dotted
lines. The small 
discontinuities in the star formation rate arise as we recompute our
model at several intervals in redshift to ensure all halos are
resolved.}
\label{fig:sfh}
\end{figure}

Figure~\ref{fig:sfh} shows the star formation rate per unit comoving
volume in our standard model ($f_{\rm esc}=100\%$) as a function of
redshift. The comparison of the solid line (final iteration) and
dashed line (penultimate iteration) shows that the model has converged
to a self-consistent star formation history in the presence of the
photoionization feedback.\footnote{We restart our calculation of the
star formation rate at several intervals in redshift to ensure all
halos are resolved. Since the merger trees used in our model do not
reproduce exactly the Press-Shechter mass function at an earlier
redshift this results in small discontinuities in the star formation
rate visible in Fig.~\ref{fig:sfh}.}  The dotted line shows for
comparison the star formation history from the model of
\scite{cole2000}, with no photoionization feedback and scaled to the
value of $\Upsilon$ used in our standard model. For $z\gsim 10$ this
is identical to that of our new model, a fact which is not surprising,
since we use the same parameters as did \scite{cole2000}, and at these
redshifts photoionization has yet to have much effect on the
IGM. (Recall that gas in halos with $T_{\rm vir} \lsim 10^4 {\rm K}$
is assumed to be unable to cool, even in the absence of an ionizing
background.) Beginning just before \HI\ reionization (as the IGM is
being reheated) the star formation rate in our model falls below that
of \scite{cole2000} as the filtering mass rises. By $z\approx 4$ the
star formation rate has recovered to the \scite{cole2000} value as the
continued formation of structure has created many halos well above the
filtering mass, and it is these which contribute most to the star
formation rate. (The filtering mass is growing only rather slowly
during this period.) The reionization of \HeII\ leads to a second
episode of reheating, leading to an increase in the filtering mass
which again suppresses star formation rates below the \scite{cole2000}
values. Once again, by $z=0$ the differences have become very small,
as star formation becomes dominated by galaxies in halos well above
the filtering mass. The effect is rather small however, with star
formation rates being reduced by around 25\% at most. The reason why
the effects of photoionization feedback on the star formation history
are quite modest in our model is that we also include supernova
feedback according to the prescription of \scite{cole2000}. This
greatly suppresses star formation in halos with circular velocities
$V_c \ll 200 {\rm km s^{-1}}$, which includes the range of halo masses
that are also affected by photoionization feedback.

For comparison, we have also computed a model in which the feedback
from supernovae is completely turned off. The other parameters in this
model are identical to those in \scite{cole2000}, apart from
$\Upsilon$, which is reduced to 0.95 to match the bright end of the
present-day galaxy luminosity function (see \S\ref{sec:fidLF}).  The
star formation rate as a function of redshift in this model, with
$f_{\rm esc}=100\%$ and photoionization feedback turned on, is shown
by the dot-dashed line in Fig.~\ref{fig:sfh}. The absence of feedback from
supernovae results in a much higher star formation rate than in our
standard model at all redshifts, but especially at high redshift,
where star formation is mostly occuring in small halos which are the
most strongly affected by supernova feedback. As a result,
reionization occurs significantly earlier in this model, at $z=11.5$
for \HI\ and at $z=4.5$ for \HeII. The short-dashed-long-dashed line
shows the star formation rate when feedback from supernovae and from
photoionization are both turned off. The reduction in the star
formation rate after reionization due to photoionization feedback is
seen to be much larger when there is no feedback from supernovae, than
in our standard model which includes supernova feedback.

\begin{figure*}
\begin{tabular}{cc}
\psfig{file=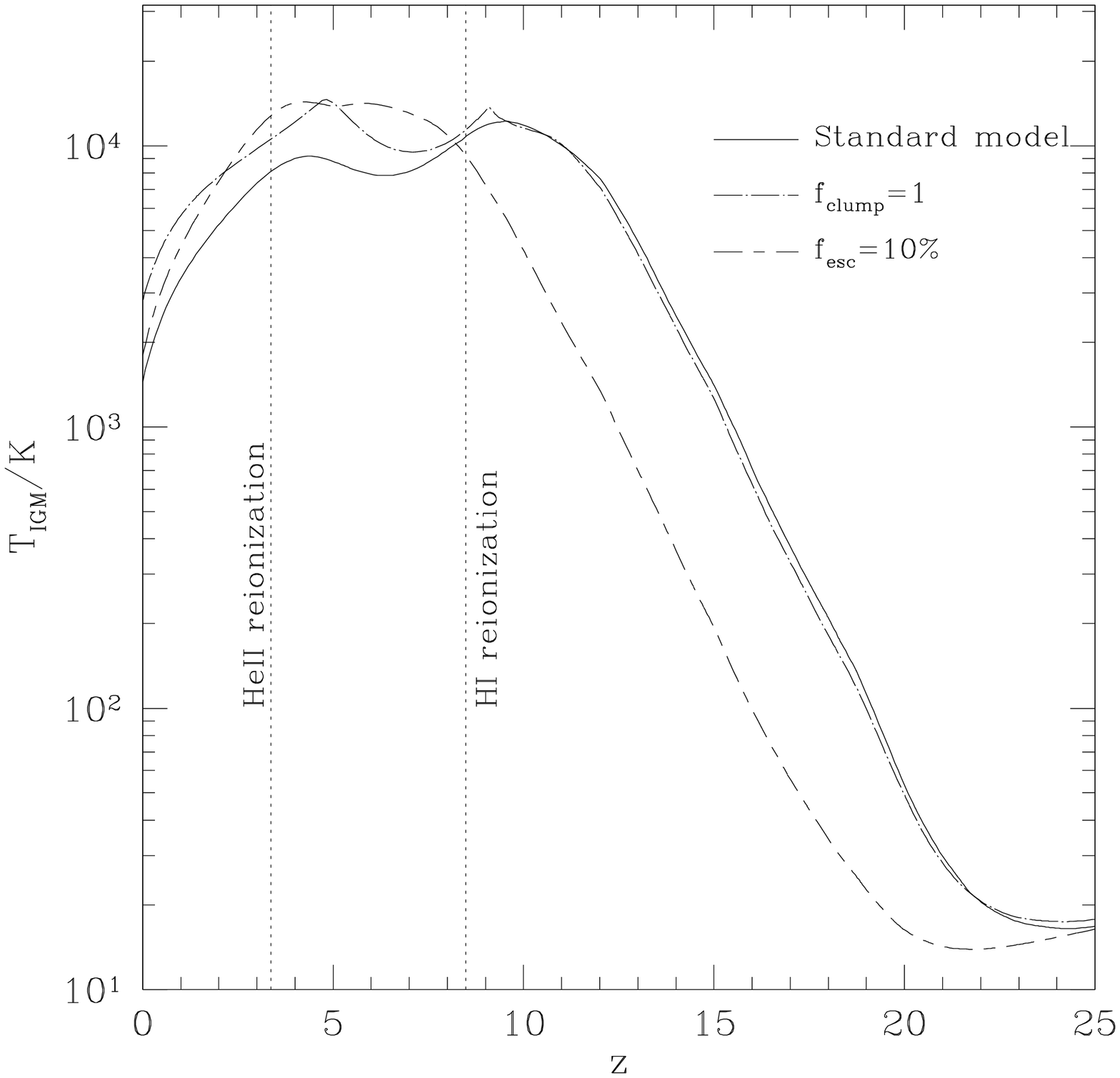,width=80mm} & \psfig{file=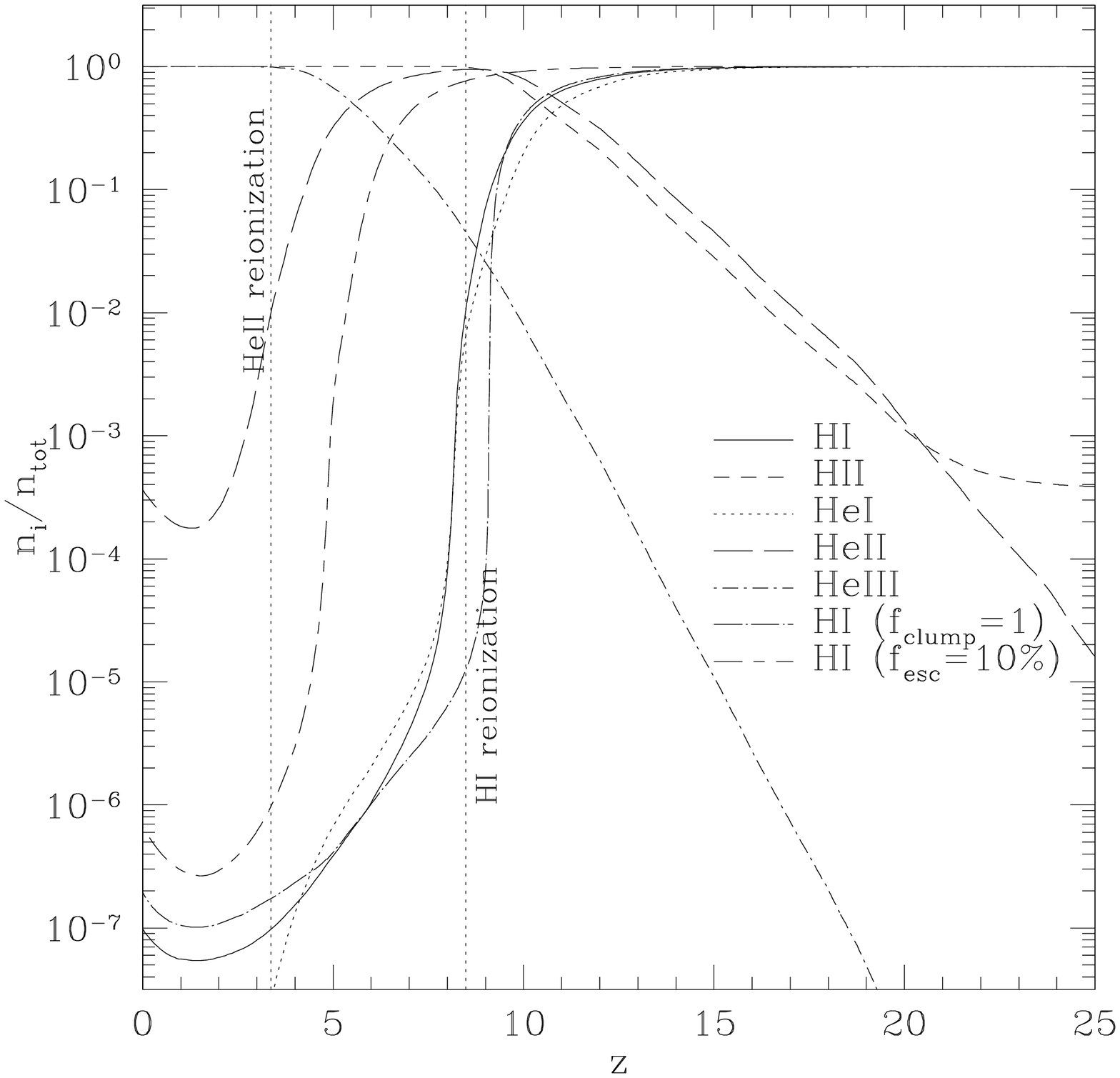,width=80mm} \\
\psfig{file=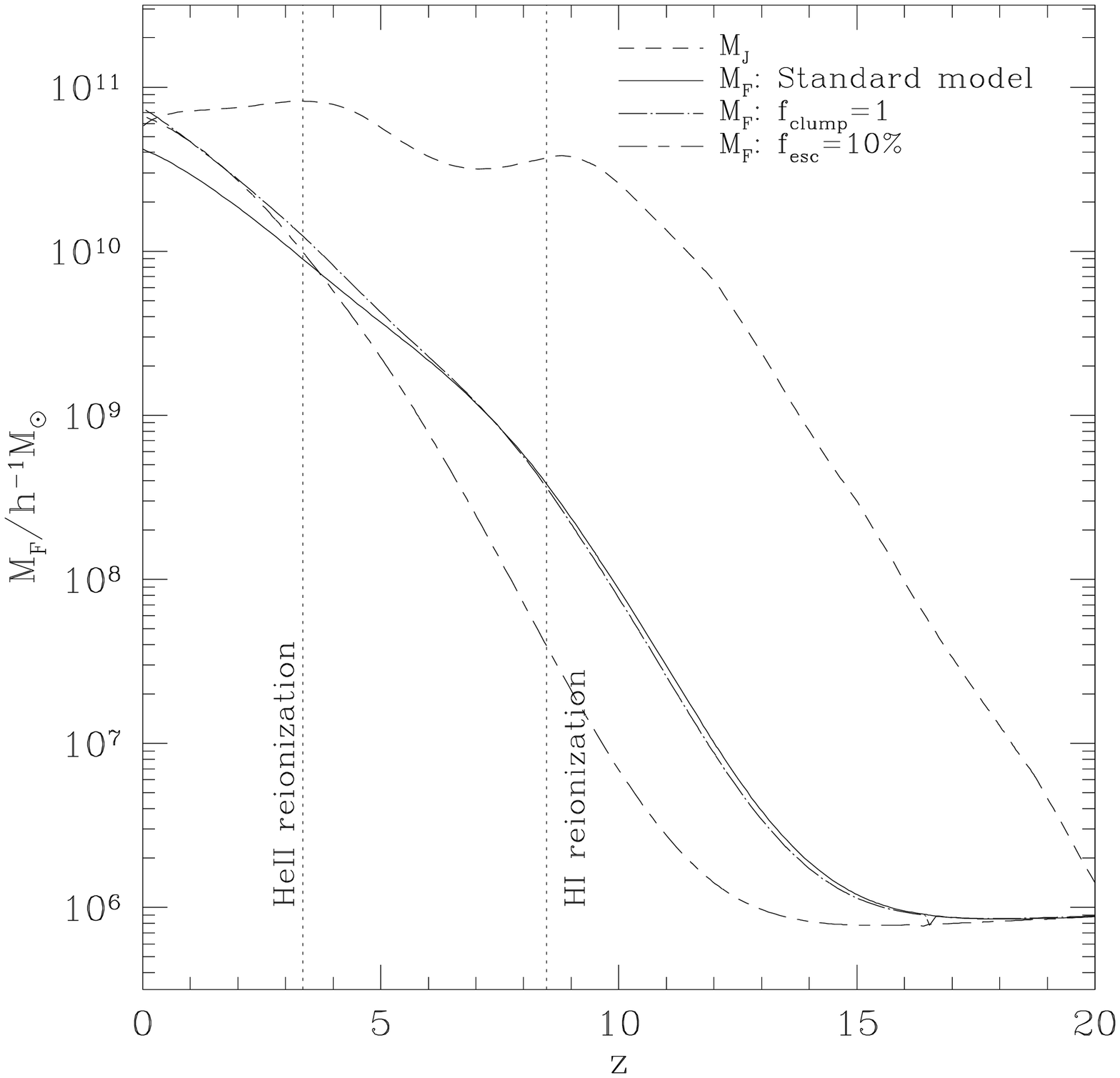,width=80mm} & \psfig{file=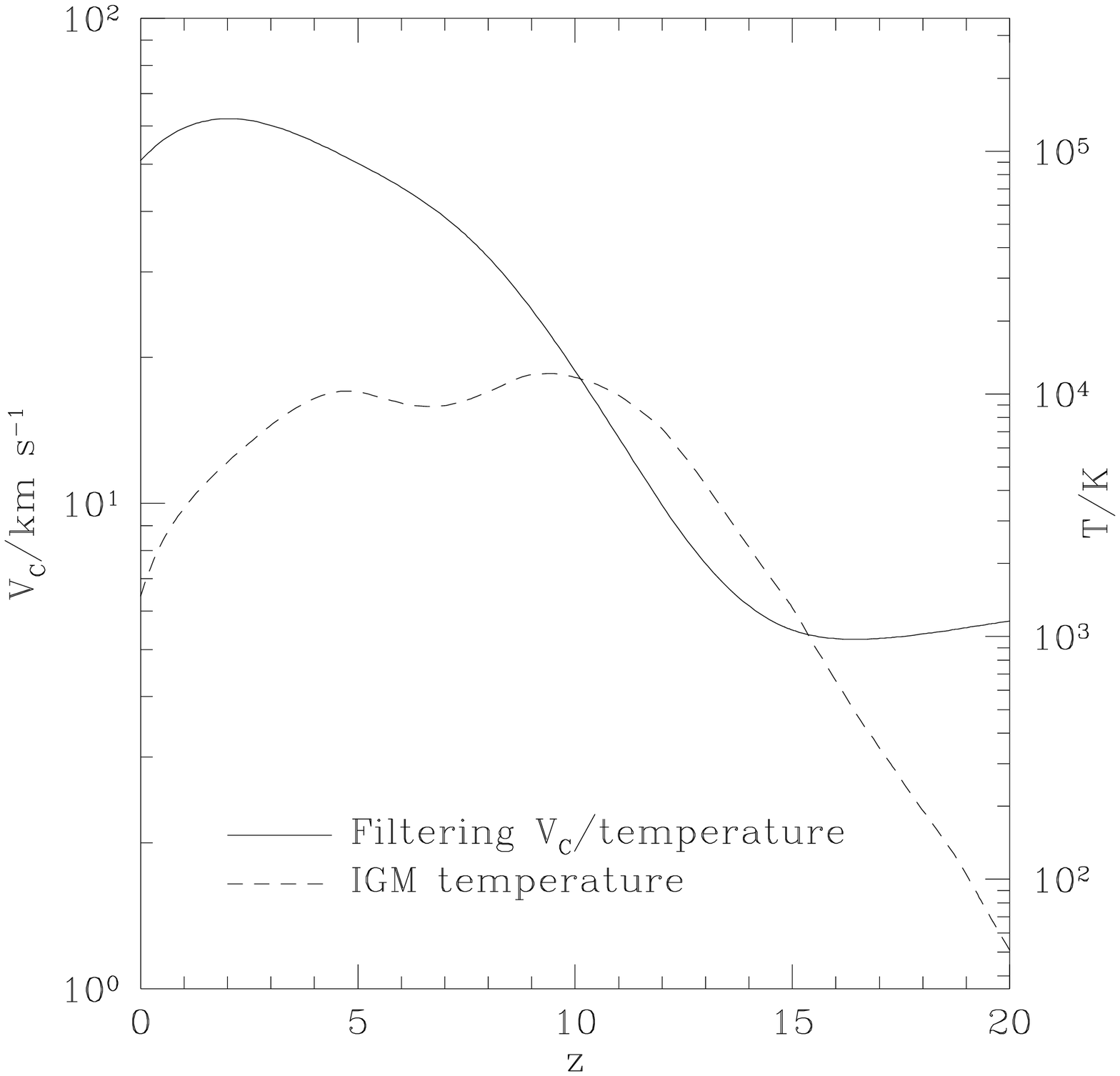,width=80mm}
\end{tabular}
\caption{\emph{Upper left-hand panel:} The volume averaged temperature
of the IGM as a function of redshift. The redshifts of reionization
for \HI\ and \HeII\ in our standard model are indicated by vertical
dotted lines. The solid line shows the standard model ($f_{\rm
esc}=100\%$), with the other lines showing the results for $f_{\rm
clump}=1$ and $f_{\rm esc}=10\%$ as indicated in the panel.  \emph{Top
right-hand panel:} The volume averaged ionization state of hydrogen
and helium as a function of redshift. The quantity shown is
$n_i/n_{\rm tot}$ where $n_{\rm tot}$ is the total abundance of the
element in question in all ionization states. The epochs of \HI\ and
\HeII\ reionization in our standard model are indicated by vertical
dotted lines (we define these as the time at which $n_i/n_{\rm tot}$
reaches 0.99). We also show $n_{\rm H{\sc i}}/n_{\rm tot}$ for models
with $f_{\rm clump}=1$ and $f_{\rm esc}=1$ as indicated in the panel.
\emph{Lower left-hand panel:} The Jeans mass (dashed line) and
filtering mass (line types as defined in the panel) as a function of
redshift. The filtering mass is shown for the standard model and also
for models with $f_{\rm clump}=1$ and $f_{\rm esc}=10\%$. The
redshifts of reionization for \HI\ and \HeII\ in our standard model
are indicated by vertical dotted lines.  \emph{Lower right-hand
panel:} The solid line shows the halo circular velocity (left-hand
axis) or halo virial temperature (right-hand axis) that correspond to
the filtering mass in the standard model. The dashed line shows the
mean IGM temperature (repeated from the top left panel).}
\label{fig:IGMfid}
\end{figure*}

\subsubsection{Thermal and Ionization History of the IGM}

Figure~\ref{fig:IGMfid} shows various properties of the IGM in our
standard model (with $f_{\rm esc}=100\%$). We also show selected
results for models with a uniform IGM (i.e. $f_{\rm clump}=1$) and for
a model with a lower escape fraction, $f_{\rm esc}=10\%$.

The top left-hand panel shows the volume averaged temperature of the
IGM as a function of redshift. At around $z=20$ ionizing photons from
stars begin to heat the IGM (at higher redshifts the temperature
scales as expected from adiabatic expansion, $T_{\rm IGM}\propto
(1+z)^2$). This results in the gas reaching a temperature of
approximately $10^4$K at $z\approx 10$ (somewhat before reionization of
\HI), at which point atomic cooling processes balance the
photoheating. The temperature then decreases until $z\approx 7$, when
the photoionization of \HeII\ by emission from quasars leads to a
second period of heating which lasts until $z\approx 5$. After this,
the gas cools rapidly until $z=0$, the cooling being due to adiabatic
expansion (due both to Hubble expansion and the expansion of the gas in
voids). The redshifts of reionization of \HI\ and \HeII\ in the
standard model are marked by vertical dotted lines. (We
define the redshift of reionization somewhat arbitrarily as the point
where 99\% of the species in question has been ionized. Since
reionization takes place rapidly the exact definition is not
important.)  With a uniform IGM, the temperature evolution is unchanged
up to just before the reionization of \HI, since the clumping in the
standard model is relatively small at these high redshifts and the
volume-weighted mean temperature shown here is dominated by the
contribution from gas close to the mean density. At lower redshifts
the clumpy IGM cools more rapidly as adiabatic expansion of gas in
voids cools the gas (and these regions are strongly weighted in the
volume-averaged temperature). The low $f_{\rm esc}$ model heats the
Universe later, as expected. There is little difference in the peak
temperature reached, which is essentially fixed by atomic physics, and
the late time temperatures are very similar to those of the standard
model.

The top-right hand panel shows the mean ionization state of hydrogen
and helium in the IGM as a function of redshift. The quantity plotted
is the average fraction of hydrogen or helium in each ionization state
(e.g. for the HI fraction $x_{\rm H{\sc i}}$, we plot $\langle n_{\rm
H{\sc ii}} \rangle 
/\langle [n_{\rm H{\sc i}}+n_{\rm H{\sc ii}}] \rangle$, where $\langle
\rangle$ denotes a volume average). Reionization is a much more rapid
process than reheating (as has been noted previously by
\scite{gnedinostriker97} and \scite{valageas99} for example). \HeI\ is
ionized by stellar photons almost simultaneously with \HI\ at a
redshift of 8, but \HeII\ is not reionized until much later ($z\approx
4$) when the harder ionizing photons from quasars become
abundant. Note that the initial decline in $x_{\rm H{\sc i}}$ is
similar in clumpy and uniform IGMs. As noted above, at these redshifts
volume-averaged quantities in the clumpy case are dominated by gas
close to the mean density, so we do not expect much difference from
the uniform case. Once started though, reionization is completed much
more rapidly in the case of a uniform IGM. In the clumpy IGM the
completion of reionization is delayed by the small fraction of high
density gas, which is reionized last. With a low escape fraction of
10\% reionization does not occur until much later, at $z\approx 5.5$
(which may be slightly too low to be consistent with recent
measurements of the Gunn-Peterson effect in quasars at $z\approx 6$:
\pcite{fan00,djorgovski01,becker01}), but otherwise proceeds in much
the same way.

The lower left-hand panel shows the evolution of the Jean's and
filtering masses with redshift. Note that both of these are defined as
halo masses, not baryonic masses. The Jean's mass simply tracks the
temperature of the IGM, while the filtering mass approximately tracks
the Jean's mass, but with a significant delay. As a result the
filtering mass can be up to 1000 times lower than the Jean's mass
during the first episode of reheating. However, at lower redshifts the
two are much more comparable and by $z=0$ the filtering mass is around
60\% of the Jean's mass. Note that the period of cooling from $z=9$ to
$z=7$ (during which time \HI\ reionization has finished, but \HeII\
reionization has yet to begin) the Jean's mass decreases slightly with
time, and as a consequence the filtering mass grows only slowly. Using
a uniform rather than a clumpy IGM affects $M_{\rm F}$ only at low
redshifts, where the lack of cool void gas in the uniform case results
in a slightly larger filtering mass. Although the filtering mass does
not begin to rise until later in a model with $f_{\rm esc}=10\%$, it
actually rises above the $f_{\rm esc}=100\%$ model at late times since
the IGM has actually been hotter in the recent past in this model.

The lower right-hand panel shows the values of the halo circular
velocity at the virial radius, and corresponding halo virial
temperature (eqn.\ref{eq:Tvir}), that correspond to the halo filtering
mass. Also shown for comparison is the average IGM temperature as a
function of redshift. It can be seen that, according to the filtering
mass prescription of \scite{gnedin2000b} that we use, the critical
halo virial temperature below which baryonic collapse into halos is
suppressed by 50\% in mass can be much greater than the IGM
temperature (by a factor 60 at $z=0$ in our standard model). In our
standard model, this temperature peaks at $T_{\rm vir} \approx 10^5
{\rm K}$, corresponding to $V_c \approx 60 {\rm km s^{-1}}$, even
though the IGM temperature is never significantly above
$10^4$K. Clearly, it will be very important to test the accuracy of
Gnedin's filtering mass prescription in greater detail using future
high resolution simulations. However, we note that similar results for
the halo circular velocity below which baryonic collapse is 50\%
suppressed were also found by \scite{quinn96}, from SPH simulations,
and \scite{thoul96}, using a 1D hydro code (they both found
$V_c\approx 50 {\rm km s^{-1}}$ at $z\approx 2$).

\begin{figure}
\psfig{file=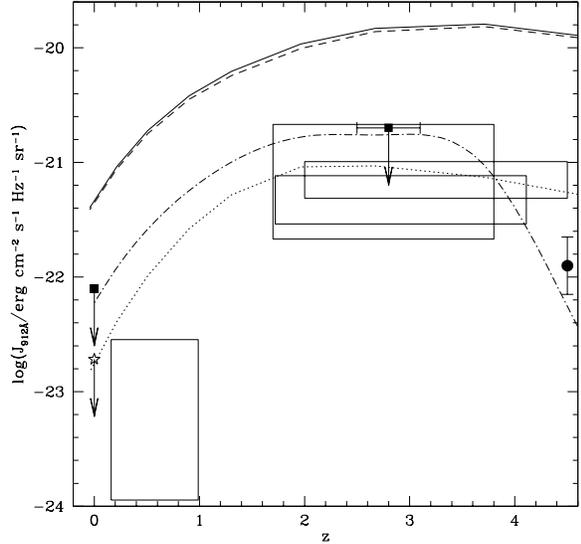,width=80mm}
\caption{Ionizing background vs redshift. The solid and dot-dashed
lines are the predicted background intensity at the Lyman limit,
$J_{\nu}(912{\rm\AA})$, for $f_{\rm esc}=100\%$ and 10\% respectively. The
dashed and dotted lines show the separate contributions from stars and
quasars in the standard model with $f_{\rm esc}=100\%$. The
rectangular boxes and the datapoint at $z=4.5$ are observational
estimates based on the proximity effect
\protect\cite{kulkarni93,BDO,scott00,cooke97,williger94}. The upper
limits are based on searches for H$\alpha$ (\protect\pcite{vogel95},
square; \protect\pcite{weyman01}, star) or Ly$\alpha$
\protect\cite{bunker98} fluorescence; the latter limit is somewhat
model-dependent.}
\label{fig:J912}
\end{figure}

\subsubsection{The Ionizing Background}

Figure~\ref{fig:J912} shows the evolution with redshift of the
predicted ionizing background at the Lyman limit, $J_{\nu}(912{\rm\AA})$.
This is compared to observational estimates from the proximity effect
in quasar spectra (e.g. \pcite{scott00}, and references therein), and
upper limits from observational searches for $H\alpha$ fluorescence
from extragalactic \HI\ clouds at low redshift \cite{vogel95} and for
$Ly\alpha$ fluorescence from Ly-limit clouds at high redshift
\cite{bunker98}. It can be seen that the background predicted for
$f_{\rm esc}=10\%$ is reasonably consistent with observational
estimates, while that predicted for $f_{\rm esc}=100\%$ is 5--10 times
too high. While this comparison seems to favour the model with $f_{\rm
esc}=10\%$, the estimated ionizing emissivity of Lyman-break galaxies
at $z\approx 3$ is only reproduced in our model with $f_{\rm
esc}=100\%$. Furthermore, with $f_{\rm esc}=10\%$, reionization seems
to occur too late compared to observational constraints from the
Gunn-Peterson effect, as already mentioned.  This contradiction in
part arises because our simple IGM model appears to underestimate the
opacity of the IGM to ionizing photons at epochs when the IGM has been
almost completely reionized. According to \scite{madau99}, at $z\lsim
5$, the Lyc opacity is dominated by the discrete absorbing clouds with
neutral hydrogen column densities $N_{HI}\sim 10^{17} {\rm cm^{-2}}$,
which produce the Lyman-limit absorption features seen in quasar
spectra. Madau et al. estimate the opacity as a function of redshift
based on the observed statistics of quasar absorption lines, and find
that the universe becomes optically thin to Lyc photons only at
$z\lsim 1.6$. In contrast, our IGM model, which lacks these absorbing
clouds, already becomes optically thin to ionizing photons at $z\sim
6$ for the case $f_{\rm esc}=100\%$. This explains why our standard
model produces a much larger ionizing background at $z\approx 3$ than
\scite{steidel00} calculate from combining their estimate of the
ionizing emissivity of Lyman-break galaxies (which our model matches)
with Madau et al.'s estimate of the Lyc opacity.

We would need to develop a much more sophisticated IGM model in order
to include the effect of discrete clouds on the Lyc opacity in a way
that was both self-consistent and agreed with observations of quasar
absorption lines. According to our models, photoionization affects
galaxy formation primarily through the effect of the IGM pressure
(which mainly depends on the redshift of reionization) rather than on
the cooling within halos (which depends on the ionizing background at
the redshift when the halo forms). Therefore, we believe that the
defficiencies of our model as regards predicting the ionizing
background after reionization should not seriously affect the
predictions that we make for galaxy formation.

\subsection{Effects on Galaxy Properties}
\label{sec:fidglob}

We now use our model, together with the properties of the IGM and
ionizing background calculated in the previous subsection, to
investigate the effects of photoionization feedback on the global
properties of galaxies at $z=0$, and to briefly consider the effects
on galaxies at higher redshifts.

\begin{figure*}
\begin{tabular}{cc}
\psfig{file=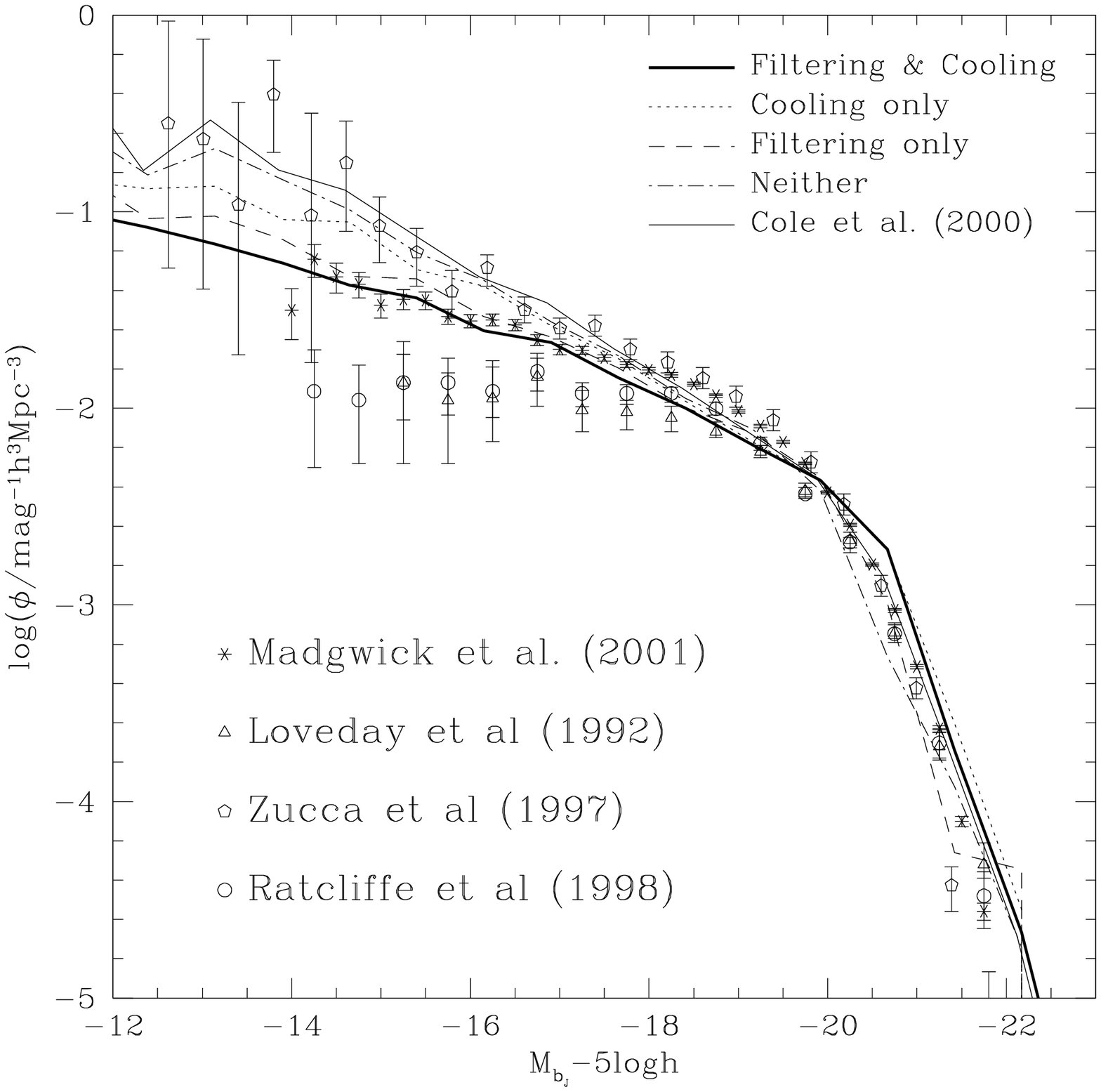,width=80mm} & \psfig{file=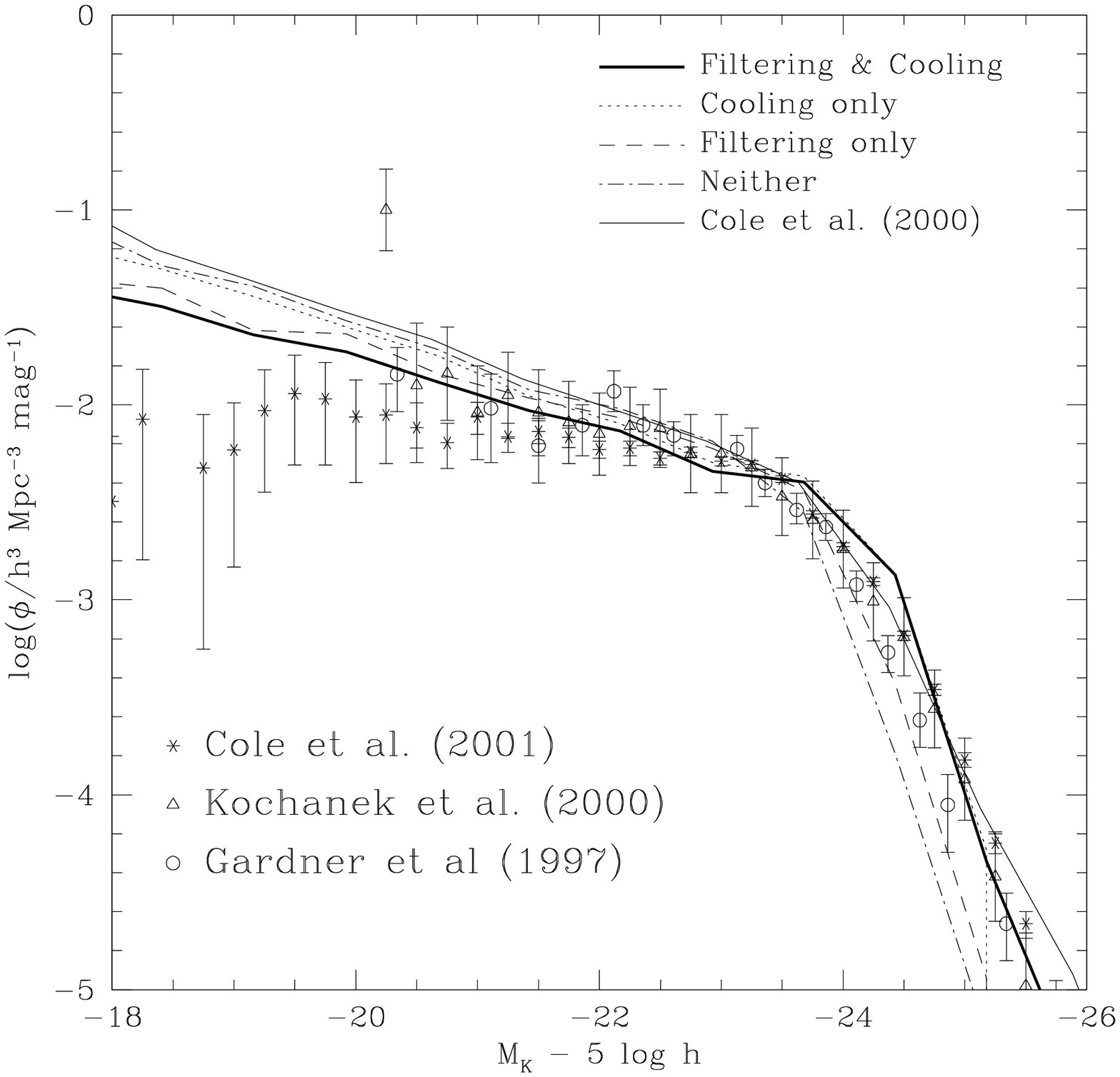,width=80mm}
\end{tabular}
\caption{Galaxy luminosity functions at z=0. The left-hand panel shows
the luminosity function in the b$_{\rm J}$-band while the right-hand
panel shows that in the K-band.  In each panel, the heavy solid line
shows the prediction of our standard model with photoionization
feedback, and the thin solid line shows the model of
\protect\pcite{cole2000}. Dotted lines show our standard model with
the effects of the filtering mass switched off, dashed lines show the
standard model with photoheating of gas in halos switched off, and
dot-dashed lines show the standard model with both of these effects
switched off.  All model luminosity functions include the effects of
dust. The symbols show observational data.}
\label{fig:fidLF}
\end{figure*}

\subsubsection{Luminosity Functions}
\label{sec:fidLF}

In Fig.~\ref{fig:fidLF} we present the B and K-band luminosity
functions for this model at z=0, and compare them to a selection of
observational data. Note that all model galaxy luminosities include
extinction by dust, calculated using the model of \scite{ferrara99} as
described by \scite{cole2000}. The heavy solid line shows the result
from our standard model, while the thin solid line shows that from the
model of \scite{cole2000}. Brighter than $L_{\star}$ the galaxy luminosity
function is mostly unaffected by the inclusion of the effects of
photoionization. Fainter than this differences become apparent, with
the luminosity function being much flatter in our present model than
in that of \scite{cole2000}. At $M_{\rm B}-5\log h\approx -13$ the
difference in amplitude of the B-band luminosity functions is about a
factor of 4. Similar behaviour is seen in the K-band. Compared to the
Cole et al. model, our new model is in appreciably better agreement
with recent observational determinations in the B and K-bands from the
2dFGRS and 2MASS galaxy surveys, by \scite{madgwick01} and
\scite{cole2mass} respectively, although in the K-band the predicted
slope is still slightly too steep.

Figure~\ref{fig:fidLF} also shows the relative importance of the new
effects we include compared to the \scite{cole2000} model. The
dot-dash line shows the effect of including tidal stripping of
satellite galaxies, but no photoionization feedback; this is seen to
reduce the number of faint galaxies very slightly. The dashed line
shows the effect of turning on the effect of IGM pressure (through the
filtering mass), but not the effect of the ionizing background on
cooling in halos, while the dotted line has the modified cooling in
halos turned on, but not the filtering mass. Comparing these, we see
for the photoionization feedback, it is primarily the effects of the
IGM pressure which suppress the galaxy formation, while the reduction
in cooling within halos has a smaller effect. (Note that these three
luminosity functions are calculated using the same value of $\Upsilon$
as for the standard model for the purposes of this comparison.)

\begin{figure}
\psfig{file=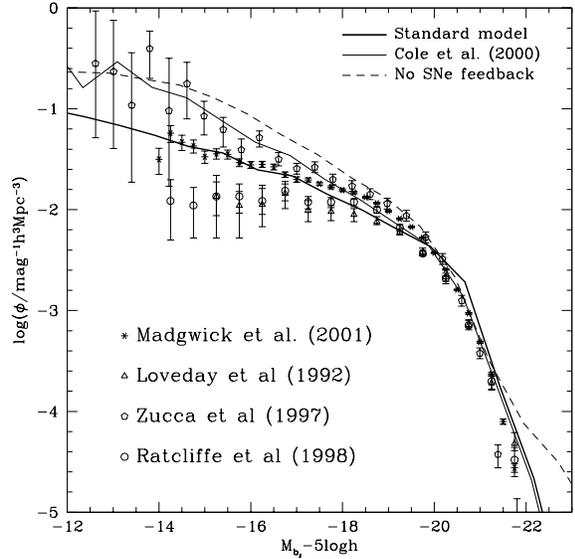,width=80mm}
\caption{The b$_{\rm J}$ band galaxy luminosity function at z=0. The
heavy solid line shows the prediction of our standard model with
photoionization feedback, and the thin solid line shows the model of
\protect\pcite{cole2000}. The dashed line shows a model with the
effects of photoionization included but without any feedback from
supernovae. All model luminosity functions include the effects of
dust. The symbols show observational data.}
\label{fig:LFBSNefree}
\end{figure}

In Fig.~\ref{fig:LFBSNefree}, we show the B-band luminosity function
in a model where photoionization feedback is included, with $f_{\rm
esc}=100\%$, but feedback from supernovae is turned off. The star
formation history for this same model was presented in
\S\ref{sec:sfrhist}. We choose $\Upsilon=0.95$ to match the amplitude
of the observed luminosity function at $L\sim L_{\star}$. The other
parameters are the same as in \scite{cole2000} and in our standard
model.  The low value of $\Upsilon$ is required because most gas which
has cooled is locked up into small objects, leaving little to form
bright galaxies. Strictly speaking, a value of $\Upsilon<1$ is
unphysical, because it requires a negative mass in brown dwarfs
(defined here as objects with $m<0.1 M_{\odot}$). However, the same
results as for $\Upsilon=0.95$ could be obtained by small
modifications to the IMF at $0.1<m< 1 M_{\odot}$, reducing the mass in
low mass stars which anyway contribute negligibly to the light from
stellar populations. This ``no SNe feedback'' model gives an
acceptable match to the observed luminosity function at the bright end
(except possibly at the highest luminosities). It predicts a faint-end
slope which is much steeper than in our standard model, but only
slightly steeper than the Cole et al. model, which had supernova
feedback but no photoionization feedback, and also only slightly
steeper than the measurement of \scite{zucca97}. The faint-end slope
is still much flatter than in a model with no feedback of any type.

We emphasize that the ``no SNe feedback'' model we have presented here
is by no means a ``best-fit'' model, since we have not varied other
parameters to achieve a better match to the luminosity function, nor
have we considered other observational constraints as Cole et
al. did. (Preliminary analysis suggests that a model with only
photoionization feedback has difficulties in matching the colours and
sizes of present-day galaxies). However, the prediction for the
faint-end slope of the luminosity function is expected to be fairly
robust, so we conclude that if the slope measured in the largest and
most recent surveys (e.g. \pcite{madgwick01}) is correct, then
photoionization feedback on its own does not produce a slope as flat
as in the real universe.  We defer a more detailed study of models
without supernova feedback to a future paper.

\begin{figure}
\psfig{file=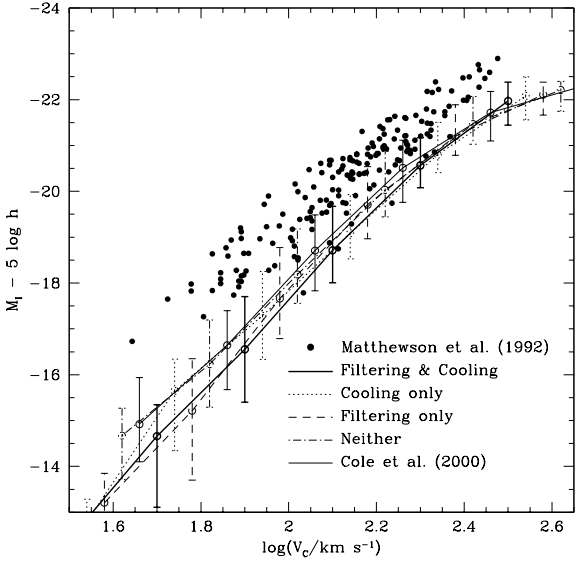,width=80mm}
\caption{The Tully-Fisher relation in the I-band at $z=0$. The lines
show the predicted median relation, with error bars indicating the
10\% and 90\% intervals of the distribution, while the filled dots
show observational data of \protect\scite{matthew92}. The heavy solid
line shows the standard model of this paper, while the thin solid line
shows the prediction from \protect\scite{cole2000}. Only star-forming
spiral galaxies are included in the model relation, with magnitudes
corrected to their face-on value including the effects of dust. The
velocities plotted are the circular velocity at the half-mass radius
of the galaxy disc. Dotted lines show our standard model with the
effects of the filtering mass switched off, dashed lines show the
standard model with photoheating of gas in halos switched off, and
dot-dashed lines show the standard model with both of these effects
switched off.}
\label{fig:TF}
\end{figure}

\subsubsection{Tully-Fisher Relation}

Figure~\ref{fig:TF} shows the I-band Tully-Fisher relation of galaxies
in our model, compared to the observational data of \scite{matthew92}.
For constructing the model relation, we select galaxies in the same
way as in \scite{cole2000}, namely we select only star-forming spiral
galaxies, but also select only those galaxies which have not been
seriously disrupted by tidal forces (specifically we remove any galaxy
which has lost more than 25\% of the mass of its disk through tidal
stripping). These strongly tidally disrupted galaxies are unlikely to
be recognisable as disks. If we do not remove these galaxies, then
Tully-Fisher relation in our model shows a scatter to very faint
magnitudes at low circular velocities, but for circular velocities
$V_c \gsim 100{\rm km s^{-1}}$ the removal of these galaxies has
little effect.

The figure also shows the model prediction of \scite{cole2000}, from
which it can be seen that the differences from our new standard model
are quite small.  We also plot lines showing the result of switching
off the effects of photoionization (as described in \S\ref{sec:fidLF}
and also in the figure caption). It can be seen from these curves that
tidal stripping makes little difference to the Tully-Fisher relation,
while the modified cooling in halos has the larger effect at high
luminosities and the IGM pressure the larger effect at low
luminosities. In any case, photoionization does not help remove the
offset in the predicted zero-point of the Tully-Fisher relation
relative to the observed one. This offset persists to bright
magnitudes, where photoionization has little effect on galaxy
formation.

\subsubsection{Further Properties of the Model at $\mathit{z=0}$}
\label{sec:fidother}

We now briefly consider the effects of photoionization feedback on
some other predicted properties of galaxies at $z=0$. We consider the
same properties as were compared with observational data in
\scite{cole2000}. Most of these comparisons concerned fairly luminous
galaxies, for which the properties in our new model are very
similar to those of the Cole et al. fiducial model, so we just
summarize the main results here.

\scite{cole2000} computed the distribution of disk scale-lengths of
spiral galaxies at different luminosities, and compared to the
observational data of \scite{dejong00}, finding good agreement in the
magnitude range they considered, $-19 > M_{\rm I} -5 \log h >
-22$. Our model produces almost identical results, as may be expected
for these bright galaxies.

Table~\ref{tb:morphs} compares the fractions of S, S0 and E galaxies
in our model brighter than $M_{\rm B}-5\log h = -19.5$
(i.e. $L_{\star}$) with \scite{cole2000} and with observational
data. We assign morphological types to our model galaxies based on
their bulge-to-total luminosity ratio in the B-band, B/T$_{\rm B}$,
(including dust extinction). Galaxies having B/T$_{\rm B}<0.4$ are
classed as S, those with B/T$_{\rm B}>0.6$ are classed as E, and those
in between are classed as S0. Our model produces a slightly higher
fraction of spirals than did that of \scite{cole2000}, a consequence
of the more detailed calculation of merger times adopted here. This is
in slightly better agreement with the observational data, but given
the crude way in which morphological types are assigned in the models,
these differences should not be over-emphasized.

\begin{table}
\begin{center}
\begin{tabular}{lc}
\hline
 & S : S0 : E \\
\hline
This work & 70 : 05 : 25 \\
\protect\scite{cole2000} & 61 : 08 : 31 \\
\protect\scite{loveday96} & 67 : 20 : 13 \\
\hline
\end{tabular}
\end{center}
\caption{The morphological mix of galaxies brighter than $M_{\rm B}-5
\log h = -19.5$ from this work and from the model of
\protect\scite{cole2000}. Also shown is the morphological mix in the
APM Bright Galaxy Catalogue (which is apparent magnitude limited) from
\protect\scite{loveday96}.}
\label{tb:morphs}
\end{table}

The cold gas content of $L_{\star}$ spiral and irregular galaxies considered
by \scite{cole2000} is unchanged in our model, as are the
metallicities of gas and stars in these galaxies. However, the
metallicity of gas in spirals and irregulars does show a somewhat
steeper dependence on luminosity than in \scite{cole2000}, resulting
in slightly better agreement with observational data. This difference
arises because of the effect of the filtering mass. In halos only
slightly more massive than the filtering mass, there can have been no
enrichment of gas in smaller halos in the merging hierarchy (as these
halos do not accrete gas). The faint central galaxies of these low
mass halos are therefore accreting relatively metal poor gas compared
to those in the \scite{cole2000} model, resulting in lower gas
metallicities. The metallicity of stars in elliptical galaxies is also
changed, but in a different way. Bright ellipticals are the same in
our model as in that of \scite{cole2000}, but faint ones on average
have somewhat higher metallicities than in Cole et al., which
worsens the agreement with the observational data. Here, the main
effect is that, with photoionization switched on, a galaxy of a given
luminosity tends to be found in a halo with higher circular velocity
(since the filtering mass reduces the amount of gas able to accrete
into each halo). The higher circular velocity implies a deeper
potential well, which makes the galaxy better at retaining metals
(i.e. fewer are lost in the winds associated with supernovae
feedback), increasing the effective yield and raising the
metallicities of the low-luminosity ellipticals. Pre-processing of gas
in lower mass halos is not so important for the ellipticals, since the
gas is processed right up to the effective yield very quickly in the
burst of star formation which makes the elliptical.

In conclusion, the largest differences in galaxy properties between
our new model and that of \scite{cole2000} occur for low luminosity
galaxies, the differences beginning to be noticeable at around 
1~magnitude faintwards of $L_{\star}$. The most important difference is a
flattening of the faint end of the galaxy luminosity function. This
slope is in reasonable agreement with the latest observational
estimates from the 2dFGRS and 2MASS galaxy surveys.

\subsubsection{Properties of Galaxies at High Redshifts}

Semi-analytic models have been used extensively to investigate the
populations of galaxies seen at high redshifts, such as Lyman-break
galaxies \cite{baugh98,governato98,somerville01}. We find that
photoionization has very little effect on the properties of
Lyman-break galaxies at $z=3$, for the range of luminosities that is
currently observed, because the filtering mass is well below the
typical mass halo in which these galaxies reside.  We defer a more
detailed consideration of high-redshift galaxies to a future paper.

\section{Discussion}
\label{sec:discuss}

We have presented a coupled model for evolution of the ionization
state and thermal properties of the IGM and the formation of
galaxies. The IGM is photoionized by radiation from stars in galaxies
and from quasars, and the photoionizing background in turn exerts a
negative feedback effect on further galaxy formation. This
photoionization feedback operates in two ways, by heating the IGM, and
so by the effects of gas pressure reducing the amount of gas which
collapses into halos, and by ionizing and heating gas within halos,
and so reducing the amount of gas able to cool to form galaxies. The
evolution of the ionizing luminosity of the galaxy population is
calculated self-consistently with the effects of this photoionization
feedback.

We calculate the formation of galaxies within the CDM model, by
adapting the semi-analytic galaxy formation model of \scite{cole2000},
modified to include the photoionization feedback effects described
above. This is coupled to a simple model for the evolution of a clumpy
IGM, which, given the evolution of the ionizing emissivity of galaxies
and quasars as an input, predicts the evolution of the mean ionized
fractions of hydrogen and helium, the volume-averaged temperature of
the IGM, and the ionizing background. We have tested the IGM model
against the results from numerical simulations of the IGM, and find
that the predictions for global properties agree reasonably well. In
particular, we find that our simple IGM model accurately predicts the
evolution of the characteristic halo mass below which accretion of
baryonic matter is strongly suppressed, which is the most important
quantity in our later study of galaxy formation.

In order to more accurately predict the properties of satellite
galaxies within larger dark matter halos, we have also improved the
\scite{cole2000} semi-analytical model by incorporating a detailed
treatment of the dynamics of satellites, including the effects of
dynamical friction, tidal stripping and heating by tidal shocks. We
have compared this model in the pure dark matter case with the results
from high-resolution N-body simulations on the amount of substructure
in dark halos, and find good agreement.  This improved model predicts
merging timescales for galaxies that on average are comparable to
those from the simple estimates used in previous work, although some
satellites have their dark halos heavily stripped by tidal forces, and
these have much longer merging timescales as a consequence of the
weaker dynamical friction force resulting from the reduced satellite
mass.

A significant uncertain parameter in our photoionization model is
$f_{\rm esc}$, the fraction of ionizing photons from stars able to
escape from galaxies. In our model, we need to assume $f_{\rm
esc}=100\%$ in order to produce the emissivity of ionizing photons at
$z=3$ inferred observationally by \scite{steidel00}. The model then
predicts reionization of \HI\ at $z\approx 8$ and reionization of
\HeII\ at $z\approx 4$. Each reionization event is preceded by an
episode of reheating. However, this model also produces an ionizing
background at $z< 4$ which is higher than observational estimates based
on the proximity effect. An escape fraction of 10\% gives much better
agreement with observational data on the ionizing background, but
produces an uncomfortably low redshift of reionization. In any case,
the choice of $f_{\rm esc}$ does not change our conclusions about the
properties of galaxies at $z=0$.

Applying our model to the evolution of the galaxy population, we find
the following results:
\begin{enumerate}
\item The global star formation rate in our model is suppressed
slightly after each episode of reheating due to reionization of \HI\
and \HeII. The suppression is quite small, with reductions of no more
than 25\% compared to a model with no reionization. By $z=0$, the star
formation rate has recovered to the level predicted by our model with
no reionization, as by then most star formation is occurring in halos
well above the masses and temperatures affected by photoionization
feedback.
\item Galaxies brighter than $L_{\star}$ are mostly unaffected by
photoionization. Faintwards of $L_{\star}$, photoionization becomes
progressively more important, reducing the abundance of galaxies of
given luminosity. Keeping the same prescription for supernovae
feedback as used by \scite{cole2000}, we find that including
photoionization feedback produces a much better fit to recent
determinations of the faint end of the galaxy luminosity function at
$z=0$ (e.g. \pcite{madgwick01,cole2mass}). Most of the effect is due
to the inability of hot IGM gas to accrete into low-mass dark matter
halos, but heating of gas in halos by the ionizing background and
tidal limitation of satellite galaxies also play a role.
\item Preliminary analysis of a model with no feedback from
supernovae, but including the effects of photoionization indicates
that such a model can produce a luminosity function with faint end
slope almost as flat as some observational estimates, and
significantly flatter than a model without supernovae feedback or
photoionization. Further work is needed to determine if such a model
can be made consistent with other observational data.
\item Other properties of bright galaxies at the present day
(e.g. sizes, Tully-Fisher relation, metallicities) are unaffected by
photoionization. For faint galaxies, we find differences in the
Tully-Fisher relation and in metallicities which are readily
understood.
\item Photoionization has little effect on the predicted properties of
Lyman-break galaxies, over the range of redshifts and luminosities for
which they are actually observed. These galaxies at $z=3$ typically
live in halos significantly more massive than that at which
photoionization feedback becomes important, so their properties are
insensitive to the reionization history.
\end{enumerate}

If the Universe was reionized through photoionization (and no
convincing alternative has been proposed), then the mechanisms
inhibiting galaxy formation which we have examined in this paper
\emph{must} operate. As such, no model of galaxy formation is complete
without their inclusion. Although we have shown that the properties of
bright galaxies are almost entirely unaffected, the properties of
faint galaxies are strongly influence by photoionization. The methods
described in this paper provide a flexible and computationally
efficient way to assess the impact of photoionization on galaxy
formation, and allow us to make definite predictions for the
properties of faint galaxies.

As we have shown, photoionization feedback has the greatest effect on
faint galaxies residing in low mass dark matter halos. As such, it
will undoubtedly have important implications for predictions about the
population of satellite galaxies found in the Local Group. In the
second paper in this series, we will explore in detail the properties
of these galaxies.

\section*{Acknowledgments}

SMC and CSF acknowledge receipt of a PPARC Advanced Fellowship and
Senior Fellowship respectively. CSF also acknowledges a Leverhulme
Research Fellowship.  CGL acknowledges support at SISSA from COFIN
funds from MURST and funds from ASI. The authors would like to thank
Nick Gnedin for supplying results from his simulations, James Taylor
for clarifying the details of his satellite dynamics calculations, and
David Weinberg for stimulating discussions. We also thank Ben Moore
for discussions on subhalo dynamics and for providing results from his
simulations.

\end{document}